\documentclass[letter]{article}

\usepackage{jheppub} 

\usepackage{graphicx,color}
\usepackage{bm}
\usepackage{amsmath}
\usepackage{amssymb}    
\usepackage{pifont}
\usepackage{stmaryrd}

\usepackage[normalem]{ulem} 
\usepackage{cancel} 
\usepackage{tikz}
\usetikzlibrary{decorations.pathmorphing}
\usetikzlibrary{decorations.markings}
\usetikzlibrary{calc}

\usepackage{standalone}
\usepackage{slashed}
\usepackage{multirow}
\usepackage{makecell}
\usepackage{tabu}
\usepackage{hhline}
\usepackage{colortbl}

\newcommand{\Ds}{\displaystyle}

\newcommand{\nn}{\nonumber}

\newcommand{\ot}{\leftarrow}

\renewcommand{\(}{\left(}
\renewcommand{\)}{\right)}
\renewcommand{\[}{\left[}
\renewcommand{\]}{\right]}

\renewcommand{\vec}[1]{\bm{#1}}

\newcommand{\blue}[1]{{\color{blue} #1}}

\newcommand{\colorTwo}[1]{{\color[rgb]{0.9,0.4,0} #1}}

\usepackage{tikz}
\usetikzlibrary{decorations.pathmorphing}
\usetikzlibrary{decorations.markings}
\usetikzlibrary{calc}

\makeatletter
\def\grd@save@target#1{%
  \def\grd@target{#1}}
\def\grd@save@start#1{%
  \def\grd@start{#1}}
\tikzset{
  grid with coordinates/.style={
    to path={%
      \pgfextra{%
        \edef\grd@@target{(\tikztotarget)}%
        \tikz@scan@one@point\grd@save@target\grd@@target\relax
        \edef\grd@@start{(\tikztostart)}%
        \tikz@scan@one@point\grd@save@start\grd@@start\relax
        \draw[minor help lines] (\tikztostart) grid (\tikztotarget);
        \draw[major help lines] (\tikztostart) grid (\tikztotarget);
        \grd@start
        \pgfmathsetmacro{\grd@xa}{\the\pgf@x/1cm}
        \pgfmathsetmacro{\grd@ya}{\the\pgf@y/1cm}
        \grd@target
        \pgfmathsetmacro{\grd@xb}{\the\pgf@x/1cm}
        \pgfmathsetmacro{\grd@yb}{\the\pgf@y/1cm}
        \pgfmathsetmacro{\grd@xc}{\grd@xa + \pgfkeysvalueof{/tikz/grid with coordinates/major step}}
        \pgfmathsetmacro{\grd@yc}{\grd@ya + \pgfkeysvalueof{/tikz/grid with coordinates/major step}}
        \foreach \x in {\grd@xa,\grd@xc,...,\grd@xb}
        \node[anchor=north] at (\x,\grd@ya) {\pgfmathprintnumber{\x}};
        \foreach \y in {\grd@ya,\grd@yc,...,\grd@yb}
        \node[anchor=east] at (\grd@xa,\y) {\pgfmathprintnumber{\y}};
      }
    }
  },
  minor help lines/.style={
    help lines,
    step=\pgfkeysvalueof{/tikz/grid with coordinates/minor step}
  },
  major help lines/.style={
    help lines,
    line width=\pgfkeysvalueof{/tikz/grid with coordinates/major line width},
    step=\pgfkeysvalueof{/tikz/grid with coordinates/major step}
  },
  grid with coordinates/.cd,
  minor step/.initial=.2,
  major step/.initial=1,
  major line width/.initial=2pt,
}

\newcommand{\drawDiagrams}{%
  \begin{tikzpicture}
  
    \coordinate (A) at (1, 1.5);
\coordinate (B) at (4, 1.5);

\draw[thick,postaction={decorate, decoration={
    markings,
    mark=at position 0.2 with {\arrow{stealth}},
    mark=at position 0.8 with {\arrow{stealth}}
    }}] ($(A)-(0,1.5)$)--($(A)+(0,1.5)$);
\draw[thick,postaction={decorate, decoration={
    markings,
    mark=at position 0.85 with {\arrow{stealth}},
    mark=at position 0.25 with {\arrow{stealth}}
    }}] ($(B)+(0,1.5)$)--($(B)+(0,-1.5)$);
\draw[thick,dashed] ($(A)+(0,1.5)$)--($(A)+(1.2,1.5)$);
\draw[thick,dashed] ($(B)+(0,1.5)$)--($(B)+(-1.2,1.5)$);

\draw[dashed,red] ($(B)+(-1.5,1.6)$)--($(B)+(-1.5,-1.6)$);

\draw[style={decorate, decoration={snake,segment length=6.2}}] ($(A)+(0,0.2)$)--($(A)+(1.5,0.)$);
\draw[style={decorate, decoration={snake,segment length=6.2}}] ($(A)+(0,-0.5)$)--($(A)+(1.5,0.)$);
\draw[style={decorate, decoration={snake,segment length=6.2}}] ($(A)+(0.3,1.5)$)--($(A)+(1.5,0.)$);
\draw[style={decorate, decoration={snake,segment length=6.2}}] ($(B)+(-0.3,1.5)$)--($(B)-(1.5,0.)$);
\draw[style={decorate, decoration={snake,segment length=6.2}}] ($(B)+(0.,-.3)$)--($(B)-(1.5,0.)$);

\draw[fill=black]  ($(A)+(0,0.2)$) circle (0.05cm);    
\draw[fill=black]  ($(A)+(0,-0.5)$) circle (0.05cm);    
\draw[fill=black]  ($(A)+(0.3,1.5)$) circle (0.05cm);    
\draw[fill=black]  ($(B)+(-0.3,1.5)$) circle (0.05cm);    
\draw[fill=black]  ($(B)+(0.,-.3)$) circle (0.05cm);    

\draw[fill=black!30]  ($(A)+(1.5,0.)$) circle (0.8cm);    

\draw[-latex,thick]($(A)+(0.2,-1.4)$)--($(A)+(0.2,-1.)$);
\node[scale=0.7] at ($(A)+(0.3,-1.6)$) {$\Ds yp_+$};
\draw[-latex,thick]($(B)+(-0.2,-1.)$)--($(B)+(-0.2,-1.4)$);
\node[scale=0.7] at ($(B)+(-0.2,-1.6)$) {$\Ds yp_+$};
\draw[-latex,thick]($(A)+(1,1.7)$)--($(A)+(2,1.7)$);
\node[scale=0.7] at ($(A)+(1.5,1.8)$) {$\Ds xp_+$};

\node[] at ($(A)+(1.5,-2.0)$) {(a)};

\coordinate (A) at (5, 1.5);
\coordinate (B) at (8, 1.5);

\draw[thick,postaction={decorate, decoration={
    markings,
    mark=at position 0.2 with {\arrow{stealth}},
    mark=at position 0.8 with {\arrow{stealth}}
    }}] ($(A)-(0,1.5)$)--($(A)+(0,1.5)$);
\draw[thick,postaction={decorate, decoration={
    markings,
    mark=at position 0.85 with {\arrow{stealth}},
    mark=at position 0.25 with {\arrow{stealth}}
    }}] ($(B)+(0,1.5)$)--($(B)+(0,-1.5)$);
\draw[thick,dashed] ($(A)+(0,1.5)$)--($(A)+(1.2,1.5)$);
\draw[thick,dashed] ($(B)+(0,1.5)$)--($(B)+(-1.2,1.5)$);

\draw[dashed,red] ($(B)+(-1.5,1.6)$)--($(B)+(-1.5,-1.6)$);

\draw[style={decorate, decoration={snake,segment length=6.2}}] ($(A)+(0,0.2)$)--($(A)+(1.5,0.)$);
\draw[style={decorate, decoration={snake,segment length=6.2}}] ($(A)+(0,-0.5)$)--($(A)+(1.5,0.)$);
\draw[style={decorate, decoration={snake,segment length=6.2}}] ($(A)+(0.3,1.5)$)--($(A)+(1.5,0.)$);
\draw[style={decorate, decoration={snake,segment length=6.2}}] ($(B)+(-0.3,1.5)$)--($(B)-(1.5,0.)$);
\draw[style={decorate, decoration={snake,segment length=6.2}}] ($(B)+(0.,-.3)$)--($(B)-(1.5,0.)$);

\draw[style={decorate, decoration={snake,segment length=6.2}}] ($(A)+(1,-1.5)$)--($(A)+(1.,0.)$);

\draw[fill=black]  ($(A)+(0,0.2)$) circle (0.05cm);    
\draw[fill=black]  ($(A)+(0,-0.5)$) circle (0.05cm);    
\draw[fill=black]  ($(A)+(0.3,1.5)$) circle (0.05cm);    
\draw[fill=black]  ($(B)+(-0.3,1.5)$) circle (0.05cm);    
\draw[fill=black]  ($(B)+(0.,-.3)$) circle (0.05cm);    

\draw[fill=black!30]  ($(A)+(1.5,0.)$) circle (0.8cm);    

\draw[-latex,thick]($(A)+(0.2,-1.4)$)--($(A)+(0.2,-1.)$);
\node[scale=0.7] at ($(A)+(0.3,-1.6)$) {$\Ds y_1p_+$};
\draw[-latex,thick] ($(A)+(1.2,-1.4)$)--($(A)+(1.2,-1.)$);
\node[scale=0.7] at ($(A)+(1.3,-1.6)$) {$\Ds y_2p_+$};
\draw[-latex,thick] ($(B)+(-0.2,-1.4)$)--($(B)+(-0.2,-1.)$);
\node[scale=0.7] at ($(B)+(-0.2,-1.6)$) {$\Ds y_3p_+$};
\draw[-latex,thick]($(A)+(1,1.7)$)--($(A)+(2,1.7)$);
\node[scale=0.7] at ($(A)+(1.5,1.8)$) {$\Ds xp_+$};

\node[] at ($(A)+(1.5,-2.0)$) {(b)};

  \end{tikzpicture}
}

\title{Transverse momentum distributions at large-$x$}

\newcommand*{\UCM}{Departamento de F\'isica Te\'orica \& IPARCOS, Universidad Complutense de Madridx, Plaza de Ciencias 1, E-28040 Madrid, Spain}
\newcommand*{\PSU}{Division of Science, Penn State University Berks, Reading, Pennsylvania 19610, USA}
\newcommand*{\JLAB}{Jefferson Lab, Newport News, VA 23606, USA}

\author{Oscar~del~Rio$^a$}
\emailAdd{oscgar03@ucm.es}
\author{Alexei~Prokudin$^{b,c}$}
\emailAdd{prokudin@jlab.org}
\author{Ignazio~Scimemi$^a$}
\emailAdd{ignazios@ucm.es}
\author{Alexey~Vladimirov$^a$}
\emailAdd{alexeyvl@ucm.es}
\affiliation{$^a$\UCM}
\affiliation{$^b$\PSU}
\affiliation{$^c$\JLAB}

\preprint{IPARCOS-UCM-25-003, JLAB-THY-25-4081}

\abstract{
We investigate the collinear matching of transverse momentum dependent (TMD) distributions at large values of $x$, computing and resumming the leading large-$x$ asymptotics for matching coefficients. The large-$x$ resummation is done directly within TMD distributions, ensuring the process-independence of the result. The derived resummation formulas are valid for all TMD distributions (except the pretzelosity). Their application improves perturbative convergence, provides practical estimation for unknown higher-order contributions, and sets restrictions for the nonperturbative part of models. Using the known anomalous dimensions, resummation can reach N$^3$LL, often exceeding the accuracy of known coefficient functions.}

\begin{document} 
\allowdisplaybreaks
\maketitle 

\section{Introduction}

The importance of resummation of the large logarithmic contributions to the perturbative series in QCD is very well known. Resummation is important from theoretical and phenomenological perspectives since it could change the asymptotic behavior and improve the perturbative convergence of the series. A particularly important example of resummation is the large-$x$, or threshold, resummation. In this case, one considers the asymptotic of Wilson coefficients functions in the regime of partons collinear momentum (Bjorken variable $x$) approaching its maximal value, $1$. In this regime coefficient functions are dominated by singular terms in the form of the ``plus" distributions $\sim [\ln^{n}(1-x)/(1-x)]_+$ and delta-functions $\delta(1-x)$. These terms can be predicted at all orders of perturbation theory and summed together resulting in a smoother behavior of the corresponding observable. Examples of large-$x$ resummation can be found in seminal papers refs.~\cite{Sterman:1986aj, Catani:1989ne}.

In this work, we study the large-$x$ resummation for Transverse Momentum dependent (TMD) distributions, see ref.~\cite{Boussarie:2023izj} for the review. We consider TMD distributions in the regime of small-$b$, with $b$ being the variable Fourier conjugated to the transverse momentum. In this regime, TMD distributions can be computed in terms of collinear distributions, such as parton distribution functions (PDFs) or fragmentation functions (FFs). The coefficient functions of small-$b$ expansion are singular in the regime $x\to1$ analogously to those of Deep Inelastic scattering or Drell-Yan process \cite{Sterman:1986aj, Catani:1989ne}. These singular terms can be predicted to all orders of perturbation theory and systematically resummed. This is not the first study of TMD and large-$x$ phenomena together. Earlier, the large-$x$ asymptotic of TMD processes was studied in the framework of the so-called joint resummation~\cite{Kulesza:2002rh, Kulesza:2003wn, Lustermans:2016nvk, Procura:2018zpn, Kang:2022nft}. In this case, one resums simultaneously large-$x$ and large-$q_T$ logarithms for the Drell-Yan process, mimicking TMD behavior in the small-$b$ regime. The conceptual difference of the present study and the joint resummation is that we consider  TMD distributions as independent and universal objects, irrespective of the process where they are observed. This perspective facilitates the applicability of our results in a wide range of applications.

The relations between TMD distributions and collinear distributions (commonly referred to as small-$b$ matching) play a crucial role in the phenomenology of TMD processes. These relations enable the usage of collinear PDFs and FFs to model the shape of TMD distributions, significantly reducing modeling freedom and enhancing the predictive power of TMD factorization. All modern determinations of TMD distributions rely on small-$b$ matching; see, for example, Refs.\cite{Moos:2023yfa, Bacchetta:2024qre, Neumann:2022lft, Billis:2024dqq} that employ N$^3$LO matching. In this respect, the large-$x$ resummation formulas derived in this work are particularly important. First, they stabilize the perturbative part of the modeling at large-$x$. Second, they provide predictions for the most significant portions of higher-loop corrections, in cases where these corrections are unknown (for example, for the helicity distribution, which is only known up to NLO~\cite{Bacchetta:2013pqa, Gutierrez-Reyes:2017glx}). Thus, the application of the large-$x$ resummation substantially enhances the predictive power of the approach. Moreover, the resummed formulas are often simpler and more practical than their fixed-order counterparts.

To avoid ambiguity in the notations of different perturbative orders, we adopt the following convention in this work. By a pure perturbative order, such as the leading order (LO), next-to-leading order (NLO), and so on, we refer to the perturbative order of the coefficient function for small-$b$ matching.  When incorporating the resummation of large-$x$ contributions, we denote the corresponding order with a subscript $x$, i.e. LL$_x$, NLL$_x$, etc. For instance, the notation NLO+N$^3$LL$_x$, indicates that the coefficient function is computed at NLO, and the large-$x$ part is resummed at N$^3$LL accuracy. This convention ensures clarity and avoids confusion with notations used for other perturbative components.

An additional important feature of our study is that we consider a wide range of TMD distributions (unpolarized, helicity, Sivers, Boer-Mulders, etc.). We explicitly demonstrate and verify that their large-$x$ asymptotics are related, but not identical. These distributions can be categorized into three universality classes based on the structure of their leading small-$b$ term:
\begin{itemize}
\item Distributions that match directly to twist-two collinear distributions. They include unpolarized, helicity, and transversity TMDPDFs and TMDFFs. Their resummation is analogous to the ordinary resummation for the Sudakov form factor \cite{Catani:1989ne}. These distributions (although indirectly) were considered within the joint-resummation approach \cite{Kulesza:2002rh, Kulesza:2003wn, Lustermans:2016nvk, Procura:2018zpn, Kang:2022nft}. Due to known expressions for the anomalous dimension, the resummation can be performed at N$^3$LL$_x$. 
\item Distributions that match to twist-two distribution at the first power in $b$. These correspond to the twist-two components of worm-gear distributions  \cite{Kanazawa:2015ajw, Scimemi:2018mmi, Rein:2022odl}.  In this case, the leading large-$x$ asymptotic is less pronounced, given by terms of the form $\sim \ln^n(1-x)$ and constants. These can be resummed in a manner similar to the previous case, up to N$^3$LL$_x$. To the best of our knowledge, our current study is the first demonstration of the resummation for the so-called Wandzura-Wilczek-type terms.
\item Distributions that match to twist-three collinear distributions. These are the Sivers, Boer-Mulders, and Collins functions, as well as the twist-three part of the worm-gear functions. In this case, we prove (and check against known computations) that the resummation is possible only up to NLL$_x$. The sub-leading terms in the asymptotic expansion contain multiparton interactions, which prevent the prediction of sub-leading terms in the resummation (at least in a straightforward manner). To our best knowledge, this constitutes the first resummation statement about the large-$x$ behavior of twist-three related observables.
\end{itemize}

Within each class, the resummation formulas are the same. The only TMD distribution not considered here is the pretzelocity TMD. This distribution corresponds to the $b^2$ component of the operator product expansion, and thus, its matching starts from a twist-four operator~\cite{Moos:2020wvd}.  Its large-$x$ resummation is left for future study.

The paper is organized as follows. In sec.~\ref{sec:main}, we present the general discussion on the structure of the small-$b$ expansion and the origin of the large-$x$ corrections. We identify the three classes of distributions presented earlier and derive the resummation for each of them in secs.~\ref{sec:tw2}, \ref{sec:WW}, and \ref{sec:tw3} correspondingly. In sec.~\ref{sec:unpol-explicitely} we outline a method to cross-check our results and directly extract the large-$x$ asymptotics from the expressions for the coefficient functions. As an example, we analyze the unpolarized TMDPDF, for which the three-loop coefficient function is known~\cite{Ebert:2020yqt, Luo:2020epw}. In sec.~\ref{sec:particular} we discuss the consequences of the large-$x$ resummation and demonstrate the numerical significance of our results. The paper is accompanied by the collection of necessary anomalous dimensions in appendix \ref{app:PT}, by the comparison with other resummation results in Mellin space in appendix \ref{app:ComparisonKang}, and by a detailed derivation of the relation between large-$N$ asymptotic in Mellin space and large-$x$ asymptotic in the momentum-fraction space, in appendix \ref{app:MellinToMomentum}. We present our conclusions in sec.~\ref{sec:conclusions}.

\section{General discussion}
\label{sec:main}

In the following we will refer to TMD  distributions in coordinate space, being $\vec{b}$ the Fourier conjugate variable of transverse momentum.
In the small-$b$ regime, $\vec b^2\ll 1/\Lambda_{\text{QCD}}^{2}$, TMD distributions can be expressed in terms of the collinear distributions (see, for instance, \cite{Becher:2010tm, Aybat:2011zv, Bacchetta:2013pqa, Echevarria:2016scs, Scimemi:2018mmi}). The small-$b$ expansion is a particular case of operator product expansion (OPE) applied to TMD operators. There are many methods to derive the coefficient functions for OPE, which are convenient in different cases. For the general discussion, as the one presented here, we found it convenient to use the terminology of background field approach (as the one used in refs.~\cite{Balitsky:1987bk, Scimemi:2019gge, Rein:2022odl}). This method makes manipulations with power-suppressed operators and equations of motion more transparent. These elements are not that important for the TMD distributions that match to twist-two directly (unpolarized, helicity, transversity distributions), where the analyses can be equally and easily done with more traditional methods (see for instance \cite{Kang:2022nft}). However, it becomes important for the TMD distributions that are proportional to $b$. Thus, we begin this section by reviewing the key steps involved in the computation of OPE and introducing the general notations used in our paper. In the subsequent subsection, we apply this framework to derive the large-$x$ resummation for specific cases of TMD distributions.

TMD distributions of the leading twist are expressed by the following matrix element
\begin{eqnarray}\label{def:TMD-gen}
\langle O_{\text{TMD}}\rangle &=& \int \frac{d\lambda}{2\pi}e^{-ix\lambda p_+}\langle p|Z_{\text{TMD}}(b;\mu,\zeta)O_{\text{TMD}}(\lambda,b)|p\rangle
\\
\nn 
&=&
\int \frac{d\lambda}{2\pi}e^{-ix\lambda p_+}\langle p|Z_{\text{TMD}}(b;\mu,\zeta)\bar q(\lambda n+b)[\lambda n+b,\pm \infty n+b]\Gamma[\pm \infty, 0]q(0)|p\rangle,
\end{eqnarray}
where $n$ is the light-cone vector $n^2=0$, $\Gamma\in\{\gamma^+,\gamma^+\gamma^5,i\sigma^{\alpha+}\gamma^5\}$ and $[a,b]$ stands for the straight Wilson line. The renormalization factor $Z_{\text{TMD}}(b;\mu,\zeta)$ removes the ultraviolet (UV) and rapidity divergences of the operator and introduces the corresponding scales $(\mu,\zeta)$. It is important to mention that the TMD renormalization factor $Z$ does not depend on $x$. Various Lorentz components of the matrix element in eq.~(\ref{def:TMD-gen}) correspond to particular TMD distributions, such as unpolarized, Sivers, helicity distribution, etc.

To compute the small-$b$ expansion, one splits the QCD fields into quantum (generically high-momentum) and background (generically low-momentum) components and perturbatively integrates them over the quantum components. The expansion in the background field is ordered in the number of background fields and has the following form:
\begin{eqnarray}
\langle O_{\text{TMD}}\rangle &=& \int d^4z_{1,2} M_2(x,b;z_{1,2}) \langle O_{2}(z_{1,2})\rangle+\int d^4z_{1,2,3} M^{\nu}_3(x,b;z_{1,2,3}) \langle O^\nu_{3}(z_{1,2,3})\rangle+...~,
\label{eq:one}
\end{eqnarray}
where $\langle O_{n}\rangle$ is a matrix element of an operator with $n$ background fields (accompanied by Wilson lines) positioned at space-time positions $z_i$, and $M_n$ is a combination of loop-integrals that appear in the integration over quantum components. The index $\nu$ is due to an extra gluon field inserted into $O_3$. The factor $Z_{\text{TMD}}$ is assigned to the functions $M$.  The expansion in eq.~\eqref{eq:one} is not yet a valid small-$b$ expansion because $M_n$'s are generic functions of $b$ and $z$'s, and are not ordered in powers of $\vec b^2$. 

The next step is to manipulate the operators $O$ and the factors $M$ to order the expansion in powers of $b$. This is achieved by expanding the background fields along the light-cone direction and integrating over the remaining components $z$. Then the fields within the operators become aligned with the light-cone direction and are accompanied by a number of off-light-cone derivatives. For example,
\begin{eqnarray}\label{OPE:derivatives}
O_{2}(z,0)=\sum_{k=0}^\infty \frac{z_T^{\mu_1}...z_T^{\mu_k}}{k!}\partial_{\mu_1}...\partial_{\mu_k}O_{2}(z^-n,0),
\end{eqnarray}
where $z_T$ is transverse or $z^+ \bar n^\mu$, and we set the second field in the operator to origin for simplicity. Since $b^\mu$ is the only transverse vector in the task, integration over $z_T$ results in a function of $b$, and thus, in massless QCD, functions $M$ become proportional to a power of $b$ (defined by the dimension of the operator) accompanied by logarithms. 

\begin{figure}[t]
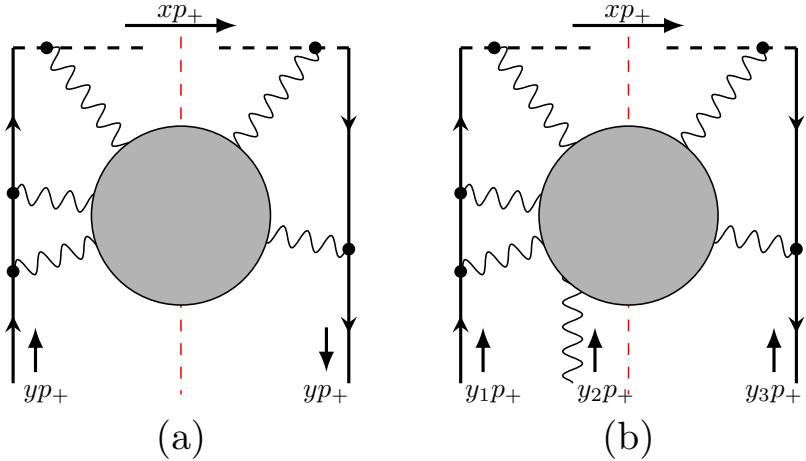

\centering
\resizebox{0.7\textwidth}{!}{\drawDiagrams}
\caption{\label{fig:diag} Diagrams contributing to the coefficient function $\widetilde{M}_2$ (a) and $\widetilde{M}_3$ (b). Only non-singlet diagrams are shown.}
\end{figure}

At this stage, one can transform the expression into momentum space. Since the parton fields are aligned along the light cone, the corresponding momenta are collinear and can be expressed via the fractions of hadron's momentum $p^+$. The resulting expansion takes the form
\begin{eqnarray}\label{OPE:via-M}
\langle O_{\text{TMD}}\rangle &=& \sum_k\int dy \widetilde{M}^{(\mu_1...\mu_k)}_2(x,y;b) \langle \widetilde{O}^{(\mu_1...\mu_k)}_{2}(y)\rangle
\\\nn && +
\sum_k\int [dy_{1,2,3}] \widetilde{M}^{(\nu,\mu_1...\mu_k)}_3(x,y_{1,2,3};b) \langle \widetilde{O}^{(\nu,\mu_1...\mu_k)}_{3}(y_{1,2,3})\rangle+...~,
\end{eqnarray}
where $k$ enumerates the number of derivatives within the operator, and $y$'s are fractions of the light-cone momenta carried by partons. Here, we imply that the matrix element is forward, and thus the sum of all parton momenta is zero. It leads to the single momentum fraction $y$ in $\widetilde{M}_2$ and to the restriction on the momentum fractions in $\widetilde{M}_3$, which is encoded in the integration measure
\begin{eqnarray}
\int [dy_{1,2,3}]=\int dy_1dy_2dy_3\delta(y_1+y_2+y_3).
\end{eqnarray}
The (bare) coefficient functions $\widetilde{M}_2$ are given by diagrams shown in fig.~\ref{fig:diag}\blue{(a)}. The coefficient function $\widetilde{M}_2$ is given by diagrams with two external legs, one carrying light-cone momentum $yp^+$ and the other $-yp^+$. The coefficient function $\widetilde{M}_3$ is given by diagrams with three external legs, fig.~\ref{fig:diag}\blue{(b)}, each carrying the light-cone momentum $y_ip^+$, $i=1,2,3$. The collinear momentum $k^+$ flowing through the operator vertex is equal to $xp^+$. The factors $z_T^{\mu}$ turn to derivatives (in momenta), and after evaluation of loop-integrals produces the powers of $b^\mu$. These rules are discussed in more detail below.

The UV and rapidity divergences of amplitudes $\widetilde{M}$ are eliminated by the factor $Z_{\text{TMD}}$ of the original TMD operator. The infra-red divergences (collinear in this case) are eliminated by the renormalization constant of the collinear operators $Z(\mu_{\text{OPE}})$. To make it explicit one could insert the unity $1=Z^{-1}(\mu_{\text{OPE}})Z(\mu_{\text{OPE}})$ into the bare expression. Then the factor $Z$ renormalizes the collinear matrix element, and $Z^{-1}$ eliminates the divergences from $\widetilde{M}$. This procedure should be done with caution, as, generally, collinear operators mix under renormalization. Therefore, before the renormalization, one must decompose operators according to a definite twist, and then group operators with the same twist together. Importantly, the procedure of twist-decomposition may turn two-point operators into three- and more-point operators due to the application of equations of motion. The final expression takes the form
\begin{eqnarray}\label{small-b}
&&\langle O_{\text{TMD}}\rangle=
C^{(0)}_{\text{tw2}}(x,b;\mu,\zeta,\mu_{\text{OPE}})\otimes \langle O_{\text{tw2}}(\mu_{\text{OPE}})\rangle
\\\nn &&
+b_\nu \[C^{(1)}_{\text{tw2}}(x,b;\mu,\zeta,\mu_{\text{OPE}})\otimes \langle O^{(\nu)}_{\text{tw2}}(\mu_{\text{OPE}})\rangle
+
C^{(1)}_{\text{tw3}}(x,b;\mu,\zeta,\mu_{\text{OPE}})\otimes \langle O^{(\nu)}_{\text{tw3}}(\mu_{\text{OPE}})\rangle\]+...~,
\end{eqnarray}
where dots denote contributions with higher powers of $b$. The coefficient functions depend on $\ln(\vec b^2)$. The sign $\otimes$ indicates the integral convolution in momentum fractions, which is generally $(n-1)$-dimensional for twist-$n$ contribution. The coefficient function of twist-2 part originates only from $\widetilde{M}_2$, the coefficient function of twist-3 part originates from $\widetilde{M}_2$ and $\widetilde{M}_3$, etc. Independent Lorentz structures of this expression are identified with particular TMD distribution (such as unpolarized, Sivers, etc.). The procedure is rather straightforward and has been applied for all TMD distributions (except the pretzelocity) at least at one-loop order \cite{Aybat:2011zv, Bacchetta:2013pqa, Scimemi:2019gge, Rein:2022odl}. The twist-2 part is particularly simple and was computed up to two- and even three-loop orders \cite{Echevarria:2015usa, Echevarria:2016scs, Gutierrez-Reyes:2018iod, Gutierrez-Reyes:2019rug, Luo:2019szz, Ebert:2020yqt, Ebert:2020qef}.

Now, we turn to the discussion of the large-$x$ limit of eq.~(\ref{small-b}). This limit has different properties for two- and three-point amplitudes and for the twist of the corresponding operator. We will discuss these cases one by one, starting from the simplest one.

\subsection{The leading term of the Operator Product Expansion}
\label{sec:tw2}

The leading term of the OPE corresponds to the zeroth power of $b$ in eq.~\eqref{small-b} and includes TMD distributions that ``survive" the naive integration over $k_T$, see ref.~\cite{Angeles-Martinez:2015sea}. These are unpolarized (both PDFs and FFs), helicity, and transversity distributions. Their coefficient functions are produced by $\widetilde{M}_2$ with collinear incoming momenta, the explicit examples of computations can be found in refs.~\cite{Aybat:2011zv, Echevarria:2011epo, Echevarria:2016scs, Gutierrez-Reyes:2017glx}. The expression for small-$b$ matching has the form of a Mellin convolution
\begin{eqnarray}\label{def:tw2-Convol}
F_{f\ot h}(x,b;\mu,\zeta)=\sum_{f'}\int_x^1 \frac{dy}{y}C_{f\ot f'}(y,b;\mu,\zeta) f_{f'\ot h}\(\frac{x}{y};\mu\)+\mathcal{O}(\vec b^2),
\end{eqnarray}
where $F$ is a TMD distribution, $f(x;\mu)$ is the corresponding collinear distribution.

The integration variable is confined to $x<y<1$, and thus at large $x$ the dominant behavior is governed by coefficient function terms that are singular at $y\to 1$. These terms are well-known and include 
\begin{eqnarray}\label{large-y-terms}
\mathcal{L}_0(y)=\delta(1-y),\quad \mathcal{L}_1(y)=\frac{1}{(1-y)_+},\quad \mathcal{L}_n(y)=\(\frac{\ln^{n-1}(1-y)}{1-y}\)_+,
\end{eqnarray}
where $n= 2,3,\dots$, and the ``plus''-distribution is defined as
\begin{eqnarray}
[f(x)]_+=f(x)-\delta(1-x)\int_0^1 dy f(y).
\end{eqnarray}
Accompanied by powers of $\alpha_s$ these terms form the series as $\alpha_s^n \mathcal{L}_n$, $\alpha_s^{n+1} \mathcal{L}_n$, etc, that could be resummed and are usually referred to as leading-logarithm (LL), next-leading-logarithm (NLL), etc, threshold resummation. By the convention of this work, we denote these orders as LL$_x$, NLL$_x$, etc.

This structure is even more evident in  Mellin space, which is defined by the transformation
\begin{eqnarray}\label{def:Mellin}
\mathbf{M}_N[F]=\int_0^1 dx x^{N-1}F(x).
\end{eqnarray}
The evaluation of $F(x)$ in the limit $x\rightarrow1$ corresponds to the evaluation of its Mellin transform $\mathbf{M}_N[F]$ in the limit $N\rightarrow\infty$. Hence, the large-$x$ asymptotic in the momentum-fraction space is equivalent to the large-$N$ asymptotic in Mellin space. The dominant terms (\ref{large-y-terms}) behave as 
\begin{equation}
    \lim_{N\to \infty}\mathbf{M}_N[\mathcal{L}_n]\sim \ln^n \bar N
\end{equation}
where $\Bar{N}=N e^{\gamma_E}$ and $\gamma_E$ is Euler's constant. Therefore computing and resumming $\alpha_s^{n+k} \ln^n N$ logarithms is equivalent to the resummation of threshold logarithms. 

As we mentioned in the Introduction, the resummation of threshold logarithms has a long history of studies \cite{Sterman:1986aj, Catani:1989ne}. Without repeating the detailed analysis, that an interested reader will be able to find in the literature, we recapitulate the final statement that at the leading power (i.e. the $\sim N^0$ part) the coefficient function is dominated by the soft-gluon radiation and reduces to the amplitude of the Wilson lines \cite{Korchemsky:1992xv}. Herewith the ``rapidity divergences'' associated with the initial parton state map onto the large-$N$ divergences \cite{Sterman:2006hu, Li:2016axz, Lustermans:2016nvk} and the ``bare" coefficient function turns to the TMD soft factor
\begin{eqnarray}
\lim_{N\to\infty}\mathbf{M}_N[C^{\text{bare}}_{f\to f'}]=\delta_{ff'}S_f^{\text{bare}}\(b,\frac{\delta}{\bar N}\),
\end{eqnarray}
where $\delta$ in the argument of $S_f^{\text{bare}}$ is the parameter that regularizes the rapidity divergences. The renormalization factor $Z_{\text{TMD}}$ is proportional to the inverse square root of the TMD soft-factor $1/\sqrt{S_f}$. Combining those together one obtains that
\begin{eqnarray}\label{MN=S}
\lim_{N\to\infty}\mathbf{M}_N[C_{f\to f'}]\equiv\lim_{N\to\infty}\mathbf{M}_N[Z_{\text{TMD}}C^{\text{bare}}_{f\to f'}]
=\delta_{ff'}\sqrt{S_f\(b;\mu,\frac{\zeta}{\bar N^2}\)},
\end{eqnarray}
where $S_f$ is the renormalized TMD soft factor for quarks or gluons, and scales $(\mu,\zeta)$ are usual scales of TMD evolution \cite{Aybat:2011zv, Scimemi:2018xaf}. The TMD soft factor, as well as, the coefficient function for the unpolarized TMD distribution are known up to three-loop order \cite{Luo:2019szz, Ebert:2020yqt, Ebert:2020qef}. Using these expressions we have explicitly confirmed the relation in eq.~(\ref{MN=S}) up to N$^3$LO order (the computation is in sec. \ref{sec:unpol-explicitely}).

The expression eq.~(\ref{MN=S}) can be written in a more practical form
\begin{eqnarray}\label{MN=e^D}
\lim_{N\to\infty}\mathbf{M}_N[C_{f\to f'}]=\delta_{ff'} \exp\(2\mathcal{D}^f_{p.t.}(b;\mu)\ln \bar N+\mathcal{E}^f(b;\mu,\zeta)\),
\end{eqnarray}
where $\mathcal{D}_{p.t.}$ is the perturbative part of the Collins-Soper kernel (also known as the so-called rapidity anomalous dimension) and $\mathcal{E}$ represents the finite at $N\to\infty$ terms. $\mathcal{E}$ coincides with the renormalized TMD soft function presented in refs.~\cite{Ebert:2020yqt, Luo:2020epw}. Both $\mathcal{D}_{p.t.}$ and $\mathcal{E}$ depend on $\ln(\vec b^2\mu^2)$. The corresponding expressions can be found in refs.~\cite{Luo:2019szz, Ebert:2020yqt}, and for convenience we have also collected them in appendix \ref{app:PT}. We have verified that our expression eq.~\eqref{MN=e^D} coincides with the result obtained in ref.~\cite{Kang:2022nft}, and we provide a detailed comparison in Appendix \ref{app:ComparisonKang}.

Phenomenological application of TMD formalism is usually performed in momentum-fraction space, therefore, we will transform eq.~(\ref{MN=e^D}) to the $x$-space. We use the relation 
\begin{eqnarray}\label{eq:MellinToMomentumFraction}
e^{\alpha \ln \bar N+\beta}=\lim_{N\to \infty}\mathbf{M}_N\[\delta(1-x)-\frac{\alpha}{(1-x)^{1+\alpha}_+}\]e^{\tilde \beta},\qquad \tilde \beta=\beta-\sum_{k=2}^{\infty} \frac{\alpha^k\zeta_k}{k},
\end{eqnarray}
where $\zeta_k$ is the Riemann zeta function, and $\alpha$ and $\beta$ are numbers. The derivation of eq.~(\ref{eq:MellinToMomentumFraction}) can be found in appendix \ref{app:MellinToMomentum}. Substituting $\alpha=2\mathcal{D}_{p.t.}$ and $\beta=\mathcal{E}$ and expanding the powers of $\alpha_s$ we confirm that this expression reproduces the terms $\sim \mathcal{L}_n$ in the small-$b$ matching coefficients up to three-loop order. We performed this comparison for unpolarized TMDPDF and TMDFF, and transversity TMDPDF.

It follows that the coefficient function for unpolarized, helicity, and transversity TMDPDFs and TMDFFs with resummed large-$x$ contributions reads
\begin{eqnarray}\label{def:Cresum}
C_{f\ot f'}(x,b;\mu,\zeta)=\delta_{ff'}V_f(x,b;\mu,\zeta)+\Delta C_{f\ot f'}(x,b;\mu,\zeta)
\end{eqnarray}
where the universal term is
\begin{eqnarray}\label{def:Vf}
V_f(x,b;\mu,\zeta)=\(\delta(1-x)-\frac{\alpha_f}{(1-x)_+^{1+\alpha_f}}\)e^{\overline{\mathcal{E}}_f}
\end{eqnarray}
with
\begin{eqnarray}\label{def:alpha-E}
\alpha_f=2\mathcal{D}^f_{p.t}(b;\mu),\qquad \overline{\mathcal{E}}_f=\mathcal{E}_f(b;\mu,\zeta)-\sum_{k=2}^{\infty} [2\mathcal{D}^f_{p.t}(b;\mu)]^k\frac{\zeta_k}{k}.
\end{eqnarray}
The part $\Delta C$ is the coefficient function at a given perturbative order with all $\mathcal{L}_n$ set to zero. It depends on a particular TMD distribution, while the term $V$ is entirely universal. Explicit expressions for perturbative series of $\mathcal{D}_{\text{pt}}$ and $\mathcal{E}$ are given in appendix \ref{app:PT}.

Equation~\eqref{def:Cresum} represents the result for large-$x$ resummation for TMD distributions that survive the naive integration over $k_T$ unpolarized, helicity, and transversity. The simplest recipe to implement eq.~\eqref{def:Cresum} in numerical codes is to remove all singular at large-$x$ contributions (delta functions and ``plus" functions) from the coefficient function (i.e. terms $\sim\mathcal{L}_n$) and add the convolution with universal $V$ term. Note, that the Mellin convolution with kernel $V$ can be expressed as
\begin{eqnarray}
\int_x^1 \frac{dy}{y}\(\delta(1-y)-\frac{\alpha}{(1-y)_+^{1+\alpha}}\)f\(\frac{x}{y}\)=
\frac{f(x)}{(1-x)^\alpha}-\alpha \int_x^1 \frac{dy}{y} \frac{f\(\frac{x}{y}\)-y f(x)}{(1-y)^{1+\alpha}}.
\end{eqnarray}

We would like to emphasize that the derived expression is valid for any set of TMD scales $(\mu,\zeta)$.

\subsection{Twist-two part for the terms linear in $b$ (Wandzura-Wilczek contribution)}
\label{sec:WW}

The terms of OPE linear in $b$ describe the matching of many TMD distributions, namely, worm-gear distributions, Sivers, Boer-Mulders, and Collins functions. These terms receive contributions from two- and three-point functions, see eq.~(\ref{OPE:via-M}). In turn, the two-point part is a mixture of contributions with twist-two (the so-called Wandzura-Wilczek (WW) part) and twist-three distributions. In this section, we discuss the WW part, while the corresponding twist-three contribution is discussed in section \ref{sec:tw3}.

The WW part contributes only to worm-gear functions \cite{Kanazawa:2015ajw, Scimemi:2018mmi}. It is produced by the twist-two part of the operator with transverse derivative $\sim \bar q \partial^\mu_T q$. Herewith, one should take into account that the operator $\bar q\partial^\mu_T q$ is not the operator of twist-two. The twist-two part of this operator is given by an integral \cite{Balitsky:1987bk}. The tree-order expressions for worm-gear functions \cite{Kanazawa:2015ajw, Scimemi:2018mmi} read
\begin{eqnarray}\label{WW-tree}
g_{1T}^{\perp,\text{WW-tree}}(x,b)=x\int_x^1 \frac{dy}{y}g_1(y),
\qquad
h_{1L}^{\perp,\text{WW-tree}}(x,b)=-x^2\int_x^1 \frac{dy}{y^2}h_1(y),
\end{eqnarray}
where $g_1$ and $h_1$ are helicity and transversity PDFs, respectively, and we omit scaling arguments for brevity.

The perturbative corrections to eq.~(\ref{WW-tree}) arise from the coefficient function $M_2^\mu$, which is given by the same diagrams as for the leading terms of OPE but with an extra factor $z_T^{\mu}$, where $z$ is the position of the background field, see eq.~(\ref{OPE:derivatives}). In momentum space, the factor $z_T^\mu$ turns to the $-i\partial/\partial p^\mu_T$, where $p$ is the momentum entering through the parton leg. It is important that the index $\mu$ is transverse, and that after the action by derivative, the incoming momentum can be turned to the light-cone momentum (like in the computation of OPE at leading power). 

It is convenient to represent the diagram with parton momentum passing along the quark line to the TMD vertex and backward to the out-going parton. In this case, there are two types of contributions with respect to the action of derivative in momentum.
\begin{itemize}
\item The contribution with the derivative acting to the $e^{i(bp)}$ in the TMD-operator vertex. This produces $b^\mu M_2$, where $M_2$ is the diagrams contributing to the leading term of OPE, shown in fig.~\ref{fig:diag}\blue{(a)}. Thus, apart from the common $b^\mu$ factor, the expression is the same as the one analyzed in the previous section. Consequently, the large-$x$ contribution of this part is also the same and equals to $b^\mu \delta_{ff'}V_f$.
\item The contribution with the derivative acting on the propagators alongside of the flow of momentum $p$. In this case, one gets the diagrams with $k^\mu_T$ in the numerator (this term can appear directly or due to the trace of gamma-matrices). Such diagrams are power-suppressed in the soft-gluon counting limit \cite{Beneke:2018gvs, Beneke:2019oqx, Bahjat-Abbas:2019fqa}, which corresponds to the large-$x$. Consequently, these diagrams do not contribute to the large-$x$ asymptotic and are vanishing at $x\to 1$.
\end{itemize}
Summarizing, the coefficient function of the WW part of TMD distributions has the same large-$x$ asymptotic as the leading contribution, thus can be presented in the form
\begin{eqnarray}\label{WW-1}
F^{\text{WW}}_{f\ot h}(x,b)=\int_x^1 \frac{dy}{y} \Big[V_f(y,b)F^{\text{WW-tree}}_{f\ot h}\(\frac{x}{y},b\)+
\sum_{f'}
\Delta C_{f\ot f'}(y,b) f_{f'\ot h}\(\frac{x}{y}\)\Big],
\end{eqnarray}
where we omit the scaling arguments for brevity and $f$ is the corresponding collinear PDF (helicity PDF for $g_{1T}^\perp$, and transversity PDF for $h_{1L}^\perp$). The coefficient function $\Delta C(y)$ is vanishing at $y\to 1$. 

To present the expressions for the WW part in the conventional form \cite{Rein:2022odl}, the order of integration in the first term of eq.~(\ref{WW-1}) should be exchanged. It results in some rather complicated expressions, however, they have very simple large-$x$ asymptotics. For distributions $g_{1T}^{\perp}$ and $h_{1L}^\perp$ we obtain
\begin{eqnarray}
&&\int_x^1 \frac{dy}{y} V_f(y,b) \frac{x}{y}\int_{x/y}^1 \frac{dz}{z}g_{1}(z)=x\int_x^1 \frac{dy}{y}\(
\frac{e^{\overline{\mathcal{E}}_f}}{(1-y)^{\alpha_f}}+...\)g_{1}\(\frac{x}{y}\),
\\
&&\int_x^1 \frac{dy}{y} V_f(y,b) \frac{-x^2}{y^2}\int_{x/y}^1 \frac{dz}{z^2}h_{1}(z)=-x^2\int_x^1 \frac{dy}{y^2}\(
\frac{e^{\overline{\mathcal{E}}_f}}{(1-y)^{\alpha_f}}+...\)h_{1}\(\frac{x}{y}\),
\end{eqnarray}
where dots denote the terms $\mathcal{O}(1-y)$. Note, that this coefficient function is less singular at large-$x$, in comparison to $\mathcal{L}_n$. It generates only a series of $\ln^n(1-x)$. Nonetheless, the rest of the terms of the coefficient function vanish at $x\to 1$.

Consequently, the expressions for small-b matching coefficients with resummed large-$x$ asymptotic for WW-part of worm-gear functions are
\begin{eqnarray}\label{WW-complete}
g_{1T}^{\perp,\text{WW}}(x,b)&=&x\int_x^1 \frac{dy}{y}\sum_{f'}\(
\delta_{ff'}V_f^{\text{WW}}\(\frac{x}{y},b;\mu,\zeta\)+\Delta C^{(g)}_{f\ot f'}\(\frac{x}{y},b;\mu,\zeta)
\)\)g_1(y),
\\
h_{1L}^{\perp,\text{WW}}(x,b)&=&-x^2\int_x^1 \frac{dy}{y^2}
\sum_{f'}\(
\delta_{ff'}V_f^{\text{WW}}\(\frac{x}{y},b;\mu,\zeta\)+
\Delta C^{(h)}_{f\ot f'}\(\frac{x}{y},b;\mu,\zeta\)\)
h_1(y),
\end{eqnarray}
where $\Delta C(x)$ are parts of coefficient function that vanish at $x\to 1$, and
\begin{eqnarray}
V_f^{\text{WW}}(x,b;\mu,\zeta)=\frac{e^{\overline{\mathcal{E}}_f}}{(1-x)^{\alpha_f}}.
\end{eqnarray}
The functions $\alpha$ and $\overline{\mathcal{E}}$ are defined in eq.~(\ref{def:alpha-E}). This result agrees with the one-loop computation of both worm-gear functions~\cite{Rein:2022odl}. 

\subsection{Twist-three part of linear-in-$b$ term}
\label{sec:tw3}

The twist-three contribution originates from two distinct sources: the two-point and three-point operators. These contributions exhibit different structures in their large-$x$ asymptotic behavior, and we analyze them separately.

The two-point contribution is treated identically to the WW part. Repeating the same steps one concludes that the large-$x$ part of the coefficient function is given by $V_f$ (\ref{def:Vf}), and the operator part is the same as at the tree-order. Therefore, the large-$x$ dominant term coming from $M_2$ has the form
\begin{eqnarray}\label{tw3-V}
F^{\text{tw3; from $M_2$}}_{f\ot f'}(x,b)=\int_x^1 \frac{dy}{y} \Big[V_f(y,b)F^{\text{tw3-tree}}_{f\ot h}\(\frac{x}{y},b\)+\mathcal{O}(1-x)\Big].
\end{eqnarray}
The tree-order expression can have different forms depending on the kind of TMD distribution. For example, for Sivers and Boer-Mulder function, $F^{\text{tw3-tree}}$ is just Qiu-Sterman projection of twist-three distributions $T(-x,0,x)$ and $E(-x,0,x)$, correspondingly. In the case of worm-gear functions $F^{\text{tw3-tree}}$ are given by the integral convolution with non-trivial kernel \cite{Rein:2022odl}. Due to it, the Sivers and Boer-Mulders functions have contributions $\sim \mathcal{L}_n$ at large-$x$, whereas in the case of worm-gear functions, the asymptotic behavior is softer.

The analysis of the three-point contribution cannot be fully addressed using the traditional methods presented in this paper and is therefore left incomplete. At present, it is not possible to express the large-$x$ asymptotics of $M_3$ in a closed form to all orders of small-$x$ resummation. The difficulty arises due to the absence of a specific kinematic limit of three-point diagram which would dominate the large-$x$ regime.  Specifically, the three-point diagrams represent the interaction of a parton pair and a single parton. Then, the large-$x$ singular terms $\sim \delta(1-x)$ (and $\sim\mathcal{L}_n$ at higher loops) are produced if a pair of partons carries its momentum to the vertex unaltered. However, it does not imply that individual momenta of partons have any specific behavior. Therefore, the large-$x$ regime does not correspond to any particular regime of the diagram. The resulting terms do not reproduce the structure of the tree-order part and do not demonstrate (at least naively) any universality. Such diagrams appear at one-loop and higher and could contribute to the large-$x$ asymptotics, at the level of NLL$_x$. In fact, it is even not clear which functional form has large-$x$ asymptotic for such diagrams, because terms produced there are not simply $\sim \mathcal{L}_n$, but involve the convolution of several variables. 

Thus, we conclude that for the twist-three part, one can resum large-$x$ asymptotics up to LL$_x$ (i.e. terms $\sim \alpha_s^n \mathcal{L}_n$, or their equivalent). New kind of contributions arise at NLL$_x$ order, and the resummation of these additional contributions requires further theoretical investigation. We leave the problem of NLL$_x$ resummation to future studies. 

Let us present an explicit check of the above statement. It can be easily done for the Sivers and Boer-Mulders functions, which are the simplest representatives of TMD distributions that match to twist-three distributions. We consider the NLO expression for small-$b$ matching computed in refs.~\cite{Scimemi:2019gge, Rein:2022odl}. They depend on several twist-three collinear distributions $T$, $\Delta T$, etc., whose precise definitions are not important here. For the definitions of twist-three distributions, we refer to the original works refs~\cite{Scimemi:2019gge, Rein:2022odl}, or to ref.~\cite{Rodini:2024usc} where all definitions are collected.

Extracting the $V$-term from the NLO expression for the Sivers function and reorganizing the remaining terms, we find
\begin{eqnarray}\label{eq:SiversResummed}
f_{1T,f}^\perp(x,b)&=&\pm \pi \frac{T_q(-x,0,x)}{(1-x)^{\alpha_f^{(1)}}}e^{\mathcal{E}_f^{(1)}}\pm \pi a_s\Bigg\{-2\mathbf{L}_\mu \tilde{\mathbb{H}}\otimes T_q(-x,0,x)+\delta \mathbf{f}_{1T}^\perp\Bigg\}+\mathcal{O}(a_s^2).
\end{eqnarray}
where $\alpha_f^{(1)}$ and $\mathcal{E}_f^{(1)}$ are the LO parts of corresponding functions. The choice of the sign $\pm$ is related to the process: the ``$+$" sign for the Drell-Yan and the ``$-$" sign for the SIDIS process. The finite part is not modified by the LL resummation and reads \cite{Scimemi:2019gge, Rein:2022odl}
\begin{eqnarray}\label{sivers:finite}
\mathbf{\delta f}_{1T}^\perp(x)&=&
\int_{-1}^1 dy \int_0^1 d\alpha \delta(x-\alpha y)
\Big[
\\\nn &&
\(C_F-\frac{C_A}{2}\)2\bar \alpha T_q(-y,0,y)+\frac{3 \alpha\bar \alpha}{2}\frac{G_+(-y,0,y)+G_-(-y,0,y)}{y}\Big].
\end{eqnarray}
The evolution kernel now becomes finite as $x\to 1$ and its action on the function $T(-x,0,x)$ is
\begin{eqnarray}\label{sivers:H}
&&\tilde{\mathbb{H}}\otimes T_q(-x,0,x)=
\colorTwo{-\(2C_F-\frac{C_A}{2}\)T_q(-x,0,x)}+
\int_{-1}^1 dy \int_0^1 d\alpha \delta(x-\alpha y)
\Bigg\{
\\\nn && 
\qquad \(C_F-\frac{C_A}{2}\)\Big[
\colorTwo{-(1+\alpha)T_q(-y,0,y)}+(2\alpha-1)_+T_q(-x,y,x-y)-\Delta T_q(-x,y,x-y)\Big]
\\\nn &&
\qquad
+\frac{C_A}{2}\Big[\colorTwo{\frac{(1+\alpha)T_q(-x,x-y,y)-2 T_q(-y,0,y)}{1-\alpha}}+\Delta T_q(-x,x-y,y)\Big]
\\\nn &&
\qquad
+\frac{1-2\alpha\bar \alpha}{4}\frac{G_+(-y,0,y)+Y_+(-y,0,y)+G_-(-y,0,y)+Y_-(-y,0,y)}{y}\Bigg\}.
\end{eqnarray}
In this expression, terms modified compared to the standard evolution equation \cite{Braun:2009mi, Rodini:2024usc} are highlighted in orange. Notably, in contrast to the complete evolution kernel, this formulation remains finite as $\alpha\to 1$ and does not incorporate any ``plus''-distributions (i.e. terms $\sim a_s \mathcal{L}_1$). However, this expression does not vanish at $x\to1$ (as it happens for twist-two contributions). It is due to $\sim a_s\mathcal{L}_0$ terms (the orange term in the first line), that are produced in the three-point diagrams, and which could not be resummed.

A similar expression applies to the Boer-Mulders function:
\begin{eqnarray}\label{eq:BoerMuldersResummed}
h_{1,q}^{\perp}(x,b;\mu,\zeta)&=&\mp \pi \frac{E_q(-x,0,x)}{(1-x)^{\alpha_f^{(1)}}}e^{\mathcal{E}_f^{(1)}}\mp \pi a_s\Big\{-2\mathbf{L}_\mu \tilde{\mathbb{H}}\otimes E_q(-x,0,x)  \Big\}+\mathcal{O}(a_s^2).
\end{eqnarray}
where the $\mp$ identifies the process: ``$-$" for DY and ``$+$" for SIDIS. The modified evolution kernel reads
\begin{eqnarray}\label{BM_k}
&& \tilde{\mathbb{H}}\otimes E_q(-x,0,x) =\colorTwo{\(\frac{3C_F}{2}-\frac{C_A}{2}\)E_q(-x,0,x)}+\int_0^1 d\alpha\int dy \delta(x-\alpha y) \Big\{
\\\nn &&\qquad
2\(C_F-\frac{C_A}{2}\)
\Big[\colorTwo{-E_q(-y,0,y)}
-\bar \alpha E_q(-x,y,x-y)\Big]
+C_A\colorTwo{\frac{E_q(-x,x-y,y)-E_q(-y,0,y)}{1-\alpha}}\Big\}.
\end{eqnarray}
Once again, modified terms are indicated in orange, relative to the complete evolution kernel. This expression also remains finite as $\alpha \to 1$, with no presence of ``plus''-distributions, but still incorporates a sub-leading large-$x$ term (the orange one in the first line). Moreover, comparing large-$x$ terms of eq.~(\ref{sivers:H}) and eq.~(\ref{BM_k}) we observe that these NLL$_x$ terms are different and, thus, non-universal. This explicitly confirms our prediction for the leading large-$x$ behavior of the twist-three part.

Note, that for both resummed expressions, eq.~\eqref{eq:SiversResummed} and eq.~\eqref{eq:BoerMuldersResummed}, we have extracted the exponential factor $e^{\mathcal{E}_f^{(1)}}$. However, the exponentiation of $\delta$-terms is not guaranteed, as some $\delta$-terms persist in the evolution kernel, as shown in eq.~\eqref{sivers:H} and eq.~\eqref{BM_k}. Nevertheless, this form simplifies the expressions considerably; for instance, the Boer-Mulders function no longer contains any non-logarithmic contributions.

We have performed the same check for the worm-gear functions $g_{1T}^\perp$ and $h_{1L}^\perp$ using the NLO expression from ref.~\cite{Rein:2022odl}, confirming the cancellation of the dominant large-$x$ terms. In these cases, the expressions are more cumbersome, involving double integrals, and are thus not presented explicitly.

\section{Explicit derivation of the large-$x$ resummation for the unpolarized TMDPDF}
\label{sec:unpol-explicitely}

In this section, we perform the check of the expression for the large-$x$ resummation eq.~(\ref{def:Cresum}) for the unpolarized TMDPDFs. Our goal is to confirm explicitly the exponentiation of the leading terms, given in eq.~\eqref{large-y-terms}, and to validate eqs.~\eqref{def:Cresum}-\eqref{def:Vf}. For this purpose we use the three-loop (N$^3$LO) coefficient functions computed in refs.~\cite{Luo:2019szz, Ebert:2020yqt, Luo:2020epw}.

In order to validate the resummation, we transform the coefficient functions to the Mellin-moment space eq.~(\ref{def:Mellin}) and consider their large-$N$ asymptotic. The coefficient functions contain three types of contributions: the plus and delta distributions eq.~\eqref{large-y-terms}, and the harmonic polylogarithms $H(a_1,\dots,a_n,x)$ \cite{Remiddi:1999ew}. The large-$N$ asymptotics of these terms are given by
\begin{align}\label{eq:MellinHPL}
&\lim_{N\rightarrow\infty}\mathbf{M}_N[H(a_1,\dots,a_n,x)]=0,
\\
\label{eq:MellinDelta}
&\lim_{N\rightarrow\infty}\mathbf{M}_N[\mathcal{L}_0(x)]=1,
\\
\label{eq:PlusDelta}
&\lim_{N\rightarrow\infty}\mathbf{M}_N[\mathcal{L}_1(x)]=-\ln\Bar{N},
\\
\label{eq:MellinLogPlusD}
&\lim_{N\rightarrow\infty}\mathbf{M}_N[\mathcal{L}_n(x)]=(-1)^{n}\Big[\frac{\ln^{n}\Bar{N}}{n}+\frac{n-1}{2}\zeta_2 \ln^{n-2}\Bar{N}
\\\nn &
\quad\quad\quad\quad\quad\quad\quad\quad\quad+\frac{(n-1)(n-2)}{2}\zeta_3 \ln^{n-3}\Bar{N}+\sum_{l=0}^{n-4}a_{nl}\ln^{l}\Bar{N}\Big],
\end{align}
where the coefficients $a_{nl}$ are combinations of powers of $\zeta_{n-l}$, see e.g. Table 1 in ref.~\cite{Catani:1989} or Section III in Ref.~\cite{Idilbi:2006dg}. Applying these transformations to the coefficient function for unpolarized TMDPDF $C_{qq}$ we confirm the exponentiation of constant and $\ln \bar{N}$ terms, in complete agreement with eq.~(\ref{MN=e^D}),
\begin{equation}\label{eq:MellinCqq}
    \lim_{N\rightarrow\infty}\mathbf{M}_N[C_{qq}]=\exp\left\lbrace2\mathcal{D}^q_{p.t.}(b;\mu)\ln\Bar{N}+\mathcal{E}^q(b;\mu,\zeta)\right\rbrace,
\end{equation}
where $\mathcal{E}^q(b;\mu)$ is the renormalized quark TMD soft function. Note, that the renormalized soft function is also computed in ref.~\cite{Ebert:2020yqt, Luo:2020epw}, and the expression for it agrees with the expression for $\mathcal{E}$.

Next, we would like to validate the resummed expression in the $x$-space and understand how the terms of the coefficient function sum into eq.~(\ref{def:Cresum}). For that purpose ,we introduce the following notations for the terms of perturbative expansion
\begin{eqnarray}\label{eq:RenormalisedOptimalSoft}
\mathcal{E}^q(b;\mu,\zeta)&=&\sum_{n=1}^\infty a^n_s(\mu) E^q_n(b;\mu,\zeta)\,,\\
\label{eq:expansionCS}
\mathcal{D}^q_{p.t.}(b;\mu)&=&\sum_{n=1}^\infty a^n_s(\mu) D^q_n(b;\mu),
\end{eqnarray}
with $a_s=g^2/(4\pi)^2$ the QCD coupling constant. The explicit expressions for coefficients $E$ and $D$ are presented in the appendix \ref{app:PT}.

We start from the leading logarithm (LL$_x$) part $\sim a_s^n \ln^n\bar N$ of eq.~(\ref{eq:MellinCqq}). It reads
\begin{equation}\label{eq:MellinLL}
\mathbf{M}_N[C^{\text{LL}_x}_{qq}]=\sum_{n=1}^\infty a^n_s(\mu)\frac{(2 D^q_1)^n}{n!}\ln^n\Bar{N},
\end{equation}
where we have omitted the arguments of the leading order (LO) Collins-Soper kernel $D^q_1(b;\mu)$ for simplicity. We can transform back this expression to $x$-space by using the first term in eq.~(\ref{eq:MellinLogPlusD}), and sum the resulting series. It gives us the LL-part of large-$x$ asymptotic for the coefficient function
\begin{equation}\label{eq:LLResumm}
C^{\text{LL}_x}_{qq}(x)=\sum_{n=0}^\infty \frac{(-2D_1^qa_s)^n}{(n-1)!}\mathcal{L}_n=\frac{-2 D^q_1a_s}{[(1-x)^{1+2 D^q_1a_s}]_+},
\end{equation}
in complete agreement with eq.~(\ref{def:Cresum}). It is interesting to mention that this order of resummation is formed solely by the DGLAP kernels because other parts of the coefficient function do not contribute to the leading asymptotic.

The other important observation is that the order of function $\mathcal{D}_{p.t.}$ must be higher by 1 compared to the order of  $\mathcal{E}$ to match the large-$x$ term. That is, as we see in this example, LL$_x$ requires $\propto a_s$ part of $\mathcal{D}_{p.t.}$ and $\propto a_s^0$ part of $\mathcal{E}$. The NLL$_x$ order requires $\propto a_s^2$ of $\mathcal{D}_{p.t.}$ and $\propto a_s$ of $\mathcal{E}$, etc. In general, N$^n$LL$_x$ requires $\propto a_s^{n+1}$ part of $\mathcal{D}_{p.t.}$ and $\propto a_s^{n}$ of $\mathcal{E}$. The present perturbative knowledge $\mathcal{D}_{p.t.}$ and $\mathcal{E}$ of allows one to reach N$^3$LL$_x$ order of large-$x$ resummation.

According to it, in order to get NLL$_x$ part we should account for terms with $D_2$ and $E_1$. This part of the series reads
\begin{equation}\label{eq:MellinNLL}
\mathbf{M}_N[C^{\text{NLL}_x}_{qq}]=\sum_{n=2}^\infty a^n_s(\mu)E_1^q\frac{(2 D^q_1)^{n-1}}{(n-1)!}\ln^{n-1}\Bar{N}+\sum_{n=2}^\infty a^n_s(\mu)2 D^q_2\frac{(2 D^q_1)^{n-2}}{(n-2)!}\ln^{n-1}\Bar{N}.
\end{equation}
Using the fact that eq.~(\ref{eq:MellinLogPlusD}) does not mix LL$_x$ series with NLL$_x$, we can transform back to $x$-space:
\begin{equation}\label{eq:NLLResumm}
C^{\text{NLL}_x}_{qq}(x)=\frac{-2a_s^2(D^q_2+E_1^qD^q_1)}{(1-x)_+^{1+2 D^q_1a_s}}+4D^q_2D^q_1a_s^3\left[\frac{\ln(1-x)}{(1-x)^{1+2 D^q_1a_s}}\right]_+.
\end{equation}
The last term in eq.~(\ref{eq:NLLResumm}) can be added into the exponent of the LL$_x$-term (\ref{eq:LLResumm}). Then, the expression for the coefficient function up to two loops is given by
\begin{eqnarray}
\nn C_{qq}(x)&=&\delta(1-x)+a_s\Tilde{C}_{qq}^{(1)}(x)+a_s^2\Tilde{C}_{qq}^{(2)}(x)\\
&&-\frac{2 D^q_1a_s}{[(1-x)^{1+2 D^q_1a_s+2D^q_2a_s^2}]_+}-\frac{2a_s^2(D^q_2+E_1^qD^q_1)}{[(1-x)^{1+2 D^q_1a_s}]_+}+\mathcal{O}(a_s^3),
\end{eqnarray}
where $\Tilde{C}_{qq}^{(n)}$ are the $n^{\text{th}}$-order coefficient functions where all plus distributions and $\delta(1-x)$-terms are omitted. 

In a similar fashion, we consider the N$^2$LL$_x$ part of the large-$x$ asymptotic. At this level, we must take into account that the second and third terms in eq.~(\ref{eq:MellinLogPlusD}) for LL and NLL parts give contributions proportional to $\zeta_2$. The resulting coefficient function reads
\begin{eqnarray}
\nn C_{qq}(x)&=&\delta(1-x)+a_s\Tilde{C}_{qq}^{(1)}(x)+a_s^2\Tilde{C}_{qq}^{(2)}(x)+a_s^3\Tilde{C}_{qq}^{(3)}(x)+a_s^4\Tilde{C}_{qq}^{(4)}(x)\\
\nn&&-2 a_s\frac{D^q_1}{[(1-x)^{1+2 D^q_1a_s+2D^q_2a_s^2+2D^q_3a_s^3+2D^q_4a_s^4}]_+}
-2a_s^2\frac{D^q_2+E_1^qD^q_1}{[(1-x)^{1+2 D^q_1a_s+2D^q_2a_s^2+2D^q_3a_s^3}]_+}\\
&&-a_s^3\,\frac{2D^q_3+2(E_1^qD^q_2+E_2^qD^q_1)+(E_1^q)^2D^q_1-4\zeta_2(D^q_1)^3}{[(1-x)^{1+2 D^q_1a_s+2D^q_2a_s^2}]_+}+\mathcal{O}(a_s^4)
\end{eqnarray}
Note, that this is not yet the final form of the resummed coefficient function because the singular terms are not yet collected into a single function.

In order to collect these terms into a single exponent, we must focus on the finite at $N\to\infty$ terms in eq.~(\ref{eq:MellinCqq}), i.e $e^{\mathcal{E}^q(b;\mu)}$. These terms generate the delta-function contributions $\sim\mathcal{L}_0(x)$ in the momentum-fraction space. Additionally, one should also take into account the constant terms that are generated by Mellin transform of $\mathcal{L}_n$ for $n>1$ (such as $\zeta_2/2$ in $\mathcal{L}_2$  eq.~(\ref{eq:MellinLogPlusD})). Then, the exponentiated expression for $\mathcal{L}_0(x)$ reads
\begin{eqnarray}
\nn C^{\delta}_{qq}(x)&=&\delta(1-x)\bigg[1-2\zeta_2a_s^2(D^q_1)^2\\
&&-4a_s^3\left(\zeta_2 D^q_1D^q_2+\frac{2}{3}\zeta_3(D^q_1)^3\right)\bigg]e^{ E_1^qa_s+ E_2^qa_s^2+ E_3^qa_s^3}\,\,+\mathcal{O}(a_s^4).
\end{eqnarray}
In total, the N$^3$LL$_x$ resumed expression up to N$^2$LO accuracy reads
\begin{eqnarray}\label{Eq:qqN3LLandN2LO}
\nn C_{qq}(x)&=&a_s\Delta{C}_{qq}^{(1)}(x)+a_s^2\Delta{C}_{qq}^{(2)}(x)+a_s^3\Delta{C}_{qq}^{(3)}(x)(x)\\
\nn &&+\delta(1-x)\left[1-2\zeta_2a_s^2(D^q_1)^2-4a_s^3\left(\zeta_2 D^q_1D^q_2+\frac{2}{3}\zeta_3(D^q_1)^3\right)\right]e^{ E_1^qa_s+ E_2^qa_s^2+ E_3^qa_s^3}\\
\nn&&-2 a_s\frac{D^q_1}{(1-x)_+^{1+2 D^q_1a_s+2D^q_2a_s^2+2D^q_3a_s^3+2D^q_4a_s^4}}
-2a_s^2\frac{D^q_2+E_1^qD^q_1}{(1-x)_+^{1+2 D^q_1a_s+2D^q_2a_s^2+2D^q_3a_s^3}}\\
\nn&&-a_s^3\,\frac{2D^q_3+2(E_1^qD^q_2+E_2^qD^q_1)+(E_1^q)^2D^q_1-4\zeta_2(D^q_1)^3}{(1-x)_+^{1+2 D^q_1a_s+2D^q_2a_s^2}}+\mathcal{O}(a_s^4)
\end{eqnarray}
where now $\Delta{C}_{qq}^{(n)}$ are the $n^{\text{th}}$-order coefficient function where both delta and plus distributions parts are omitted. This expression can be simplified even further if we extract the same coefficient out of the plus-distribution term
\begin{eqnarray}\label{Eq:qqfinalresumm}
\nn C_{qq}(x)&=&a_s\Delta{C}_{qq}^{(1)}(x)+a_s^2\Delta{C}_{qq}^{(2)}(x)+a_s^3\Delta{C}_{qq}^{(3)}(x)(x)\\
&&+\left[\delta(1-x)-\frac{2 D^q_1a_s+2D^q_2a_s^2+2D^q_3a_s^3}{(1-x)_+^{1+2 D^q_1a_s+2D^q_2a_s^2+2D^q_3a_s^3}}\right]e^{E_1^qa_s+ E_2^qa_s^2+ E_3^qa_s^3}\\
\nn&&\times e^{-2 (2 D^q_1a_s+2D^q_2a_s^2)^2\zeta_2-\frac{8}{3} (2 D^q_1a_s+2D^q_2a_s^2)^3\zeta_3}\,\,+\mathcal{O}(a_s^4)\\
\nn&&=\Delta{C}_{qq}(x)+\left[\delta(1-x)-\frac{2\mathcal{D}^q_{p.t.}(b;\mu)}{(1-x)_+^{1+2\mathcal{D}^q_{p.t.}(b;\mu)}}\right]e^{\mathcal{E}^q(b;\mu)-\sum_{k=2}^\infty [2\mathcal{D}^q_{p.t.}(b;\mu)]^k\frac{\zeta_k}{k}}
\end{eqnarray}
This result coincides with the one previously presented in  eqs.~\eqref{def:Cresum}-\eqref{def:Vf}. 

In this way, we have explicitly confirmed our formula eq.~(\ref{def:Cresum}) for the large-$x$ resummation. We have performed the same check for the gluon-gluon part of the coefficient function. In this case, all anomalous dimensions must be replaced by their gluon analogs. The off-diagonal-flavor parts of the coefficient functions do not incorporate $\mathcal{L}_n$ distributions, and therefore, do not contribute to the large-$x$ asymptotic. This also agrees with eq.~\eqref{def:Cresum}, where $\delta_{ff'}$ ensures that $C_{qg}=\Delta C_{qg}$ and $C_{gq}=\Delta C_{gq}$. 

We have also performed the same manipulations for unpolarized TMDFF. The expressions for the coefficient functions are given in ref.~\cite{Luo:2020epw}. We have confirmed that the large-$x$ resummation for the TMDFF is identical to those of TMDPDF with the replacement $x\rightarrow z$.

\section{Phenomenological results}
\label{sec:particular}

In this section, we discuss the implications of the large-$x$ resummation for the phenomenology of TMD distributions. The resummation can be easily implemented in the existing codes for TMD phenomenology. In particular, we have added this option into \texttt{artemide} \cite{artemide}. Generally, the usage of large-$x$ resummation helps to saturate the perturbative series at large-$x$, $x\gtrsim 0.2$. However, numerically the effect is not very large: of the order of 10\% in comparison to NLO, and smaller for higher orders. More importantly, the usage of the large-$x$ resummation sets restrictions on the modeling of TMD distribution, namely, it restricts the constraints of the model for $b^*$, as discussed below.

\subsection{Restriction on the models of TMD distributions}

TMD distributions are non-perturbative functions of $x$ and $b$. The small-$b$ matching, discussed in the previous section, must be incorporated into the ansatz, but should not bias the non-perturbative part. The usual way to parameterize the fitting ansatz for TMD distribution $F$ has the following structure
\begin{eqnarray}\label{ansatz-1}
F(x,b;\mu,\zeta)=C(x,b^*(b);\mu,\zeta,\mu_{\text{OPE}})\otimes f(x;\mu_{\text{OPE}}) \cdot f_{\text{NP}}(x,b),
\end{eqnarray}
where $C\otimes f$ is the integral convolution representing the small-$b$ matching (with $f$ being some collinear distribution). The function $f_{\text{NP}}$ is a function to fit with the only restriction that $f_{\text{NP}}(b)= 1+\mathcal{O}(b^2)$ at $b\to 0$. The function $b^*(b)$ must be $b^*(b)\sim b$ at small $b$. This ansatz preserves the small-$b$ behavior while leaving sufficient freedom to parametrize the non-perturbative part.

In expression (\ref{ansatz-1}) we have especially indicated the scales of QCD that enter in various parts of the ansatz. The understanding of these scales is important for the implementation of the large-$x$ resummation. The scales $\mu$ and $\zeta$ are set by the kinematics of the process, and per se do not affect the structure and properties of the OPE. The terms containing the logarithms of these scales can be collected into a universal factor (such as $\exp(\overline{\mathcal{E}})$ in eq.~(\ref{def:Vf})) and moved outside of the integral convolution. The scale of the operator product expansion is denoted by $\mu_{\text{OPE}}$. Often, the scales $\mu$ and $\mu_{\text{OPE}}$ (and even $\zeta$) are equalized, which complicates the theoretical analysis since in this case one mixes the effects of the evolution and non-perturbative motion of partons. The more refined procedure, such as $\zeta$-prescription~\cite{Scimemi:2018xaf,Scimemi:2019cmh}, separates the evolution scales.

The scale of OPE, $\mu_{\text{OPE}}$, enters the coefficient function as argument of $\alpha_s(\mu_{\text{OPE}})$ and the argument of the logarithms
\begin{eqnarray}
\mathbf{L}_\mu=\ln\(\frac{b_*^2(b)\mu^2_{\text{OPE}}}{4e^{-2\gamma_E}}\).
\end{eqnarray}
There are no obvious limitations on the value of $\mu_{\text{OPE}}$. The dependence on $\mu_{\text{OPE}}$ cancels in between the coefficient function and the PDF evolution. However, there are soft limitations which can be summarized as
\begin{eqnarray}\label{restriction-1}
\mu_{\text{OPE}}\gg \Lambda_{\text{QCD}},\qquad \lim_{b\to 0}\mathbf{L}_\mu \lesssim 1.
\end{eqnarray}
The first requirement is needed to ensure the existence of the perturbative expansion, while the second is required to guarantee the numerical stability of the coefficient function in the region of the validity of perturbative expansion. Generally, the expression for $b^*$ remains unrestricted because for $b\gtrsim 1 $ GeV$^{-1}$ the non-perturbative effects dominate the OPE and the possible logarithmic growth of the coefficient function in eq.~(\ref{ansatz-1}) is tamed by the non-perturbative function.

The situation however changes when one implements the large-$x$ resummation for the coefficent function. The resummed expression eq.~(\ref{def:Vf}) depends of $\alpha_f$, which in turn depends of $\mu_{\text{OPE}}$ and $\mathbf{L}_\mu$. This function must satisfy 
\begin{eqnarray}\label{restriction-alpha}
\alpha_f(b^*(b);\mu_{\text{OPE}})<1,
\end{eqnarray}
strictly, in the full range of $b$. If this restriction is violated then the integral convolution diverges. To facilitate the convergence and stability of the ansatz one can set $\alpha_f\ll 1$.

The function $\alpha$ is given by the rapidity anomalous dimension (\ref{def:alpha-E}), which has the following structure
\begin{eqnarray}
\mathcal{D}^f_{p.t.}(b;\mu)=\sum_{n=1}^\infty \sum_{k=0}^{n} C_fa_s^n(\mu) \mathbf{L}_\mu^k d_{n,k}.
\end{eqnarray}
The coefficient $d_{1,0}$ is equal to zero, and the LO expression for rapidity anomalous dimension for quark is simply
\begin{eqnarray}
\mathcal{D}^q_{p.t.}(b;\mu)=2C_Fa_s(\mu)\mathbf{L}_\mu+\mathcal{O}(\alpha_s^2),
\end{eqnarray}
where $C_F=4/3$ (the Casimir eigenvalue for the fundamental representation of $SU(3)$, it should be replaced by the Casimir operator of the adjoint representation of  $SU(3)$, $C_A=3$, for the gluon case). Assuming that $a_s\ll 1$, we must restrict $\mathbf{L}_\mu\lesssim 1$, in the full range of $b$. Therefore, to guarantee the convergence of the integral convolution with large-$x$ resumed function, we must request that restriction in eq.~(\ref{restriction-1}) is valid in the full range of $b$.

Importantly, the rapidity anomalous dimension for the gluon is approximately $C_A/C_F = 9/4$ times larger than the quark anomalous dimension due to the Casimir scaling that is valid for the first three terms of the perturbative expansion. Therefore, the restrictions for the $b^*$ are much more important in the case of gluon distributions.

One can roughly estimate the maximum allowed value of $b^*(b)$ using the LO solution for $a_s$, $a_s^{-1}=\beta_0 \ln(\mu^2/\Lambda^2)$. Then the restriction in eq.~(\ref{restriction-alpha}) roughly gives
\begin{eqnarray}
b^*\lesssim \Lambda_{\text{QCD}}^{-\frac{\beta_0}{4C_F}}\approx 5.5 \; \text{GeV}^{-1}~\text{(for the quark)},
\qquad
b^*\lesssim \Lambda_{\text{QCD}}^{-\frac{\beta_0}{4C_A}}\approx 2\; \text{GeV}^{-1}~\text{(for the gluon)},
\end{eqnarray}
where we use $\Lambda_{\text{QCD}}\approx 300$ MeV. This is only a rough estimate, presented here in order to understand the size of the effect. In phenomenological applications one should construct $b^*$ using the actual values of $\mu_{\text{OPE}}$ and $a_s(\mu)$.

\begin{figure}
\centering
\includegraphics[width=0.48\linewidth]{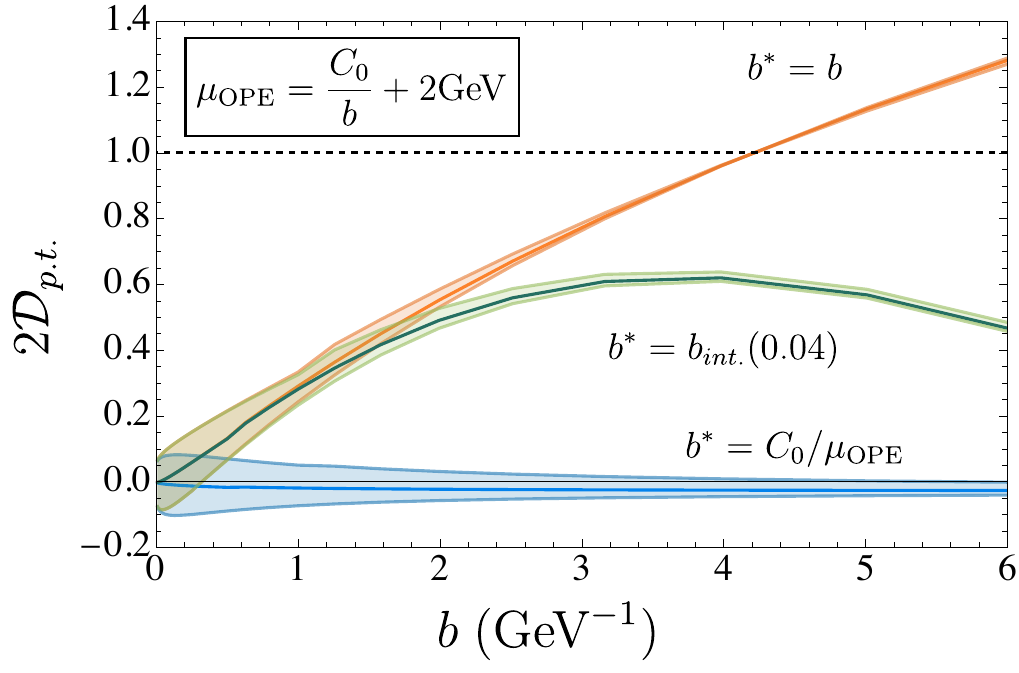}
\includegraphics[width=0.48\linewidth]{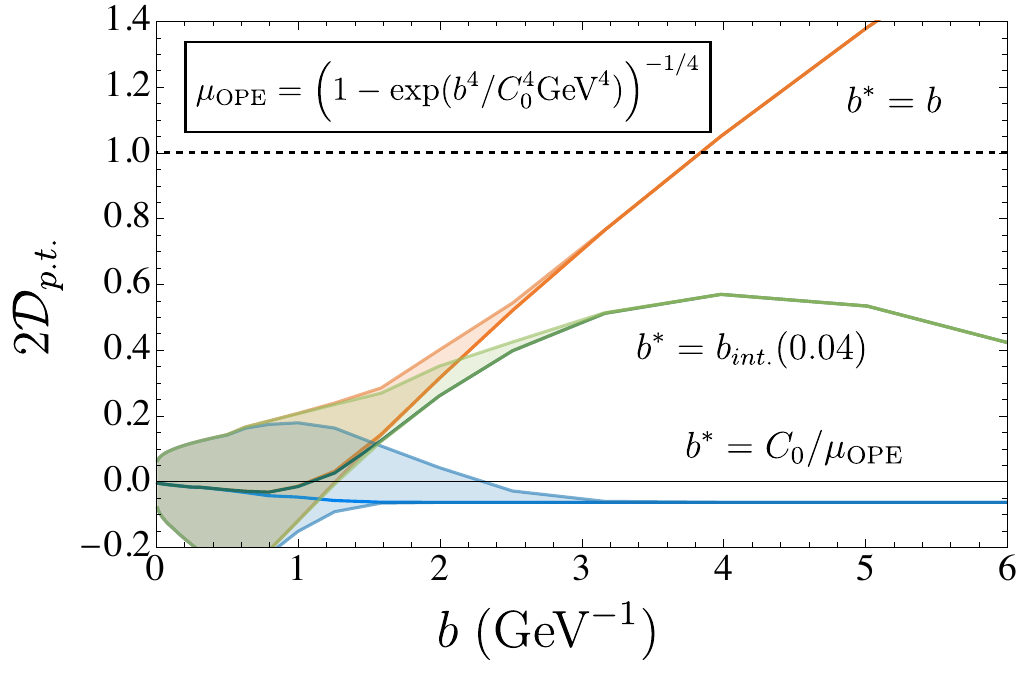}
\caption{
\label{fig:alpha} Behavior of $\alpha_f=2\mathcal{D}_{p.t.}$ as a function of $b$ in different setups of $\mu_{\text{OPE}}$ and $b^*$. The left (right) panel is for $\mu_{\text{OPE}}$ used in ART23 (MAP24, without $b_{\text{min}}$) fits. The different curves correspond to different values of $b^*$, which are indicated in the plot. The bands are obtained by variation of $\mu_{\text{OPE}}\to c_4\,\mu_{\text{OPE}} $ with $c_4\in[0.5,2.]$.}
\end{figure}

Theoretically, the choice of $\mu_{\text{OPE}}$ should not impact the values of TMD distributions. However, practically, it influences the shape of TMD distribution at large and intermediate $b$'s, and thus is a constituent part of the non-perturbative modeling. The same holds for $b^*$ function, as far as, $b^*\simeq b$ at small-$b$. One can find a lot of different implementations of $\mu_{\text{OPE}}$ and $b^*$ in the literature, compare for instance \cite{Scimemi:2019mlf, Bacchetta:2019sam, Moos:2023yfa, Bacchetta:2024qre, Billis:2024dqq, Camarda:2019zyx, Camarda:2023dqn, Neumann:2021zkb, Piloneta:2024aac}. These are typical cases of $b^*$: 
\begin{itemize}
\item The \text{local}-$b^*$ with $b^*\propto 1/\mu_{\text{OPE}}$. In this case $\mathbf{L}_\mu=0$ in the whole range of $b$, and thus the function $\alpha_f<0$ at all values of $b$. In this way, local-$b^*$ ansatz automatically satisfies the restrictions posed by the large-$x$ resummation. However, it leads to a somewhat poorer description of the low-energy data.
\item The \text{global}-$b^*$ with $b^*=b$. In this case $\mathbf{L}_\mu$ grows as $\ln(b^2)$ and eventually $\alpha_f$ became bigger than $1$. Therefore, it cannot be used together with large-$x$ resummation, although it nicely fits the data (see for instance ART23 model \cite{Moos:2023yfa}).
\end{itemize}
To fix the problem of growing $b^*$, and simultaneously preserve the predictive power of the ART23 model, we suggest the following interpolated form
\begin{eqnarray}
b_{int.}(b)=b e^{-a b^2}+\frac{C_0}{\mu_{\text{OPE}}(b)}(1-e^{-ab^2}),
\end{eqnarray}
with $a=0.04$ GeV$^2$ ($\approx \Lambda^2_{\text{QCD}}$). With this choice, and a minimal tune of non-perturbative parameters, the predictive power of the ART23 model remains the same as in the original form \cite{Moos:2023yfa}. In fig.~\ref{fig:alpha}, we compare the behavior of $\alpha_f$ as the function of $b$ for $\mu_{\text{OPE}}$ from ART23 \cite{Moos:2023yfa} and MAP24 \cite{Bacchetta:2024qre} (without $b_{\text{min}}$ part), for different choices of $b^*$. The band demonstrate the scale-variation uncertainty for variation of $\mu_{\text{OPE}}\to \mu_{\text{OPE}} c_4$ function with parameter $c_4\in[0.5,2.]$ (the variation affects only the small-$b$ part). The plots are made at $\alpha_s^4$ perturbative accuracy (N$^3$LO).

\subsection{The unpolarized TMDPDF}

\begin{figure}
\centering
\includegraphics[width=0.49\linewidth]{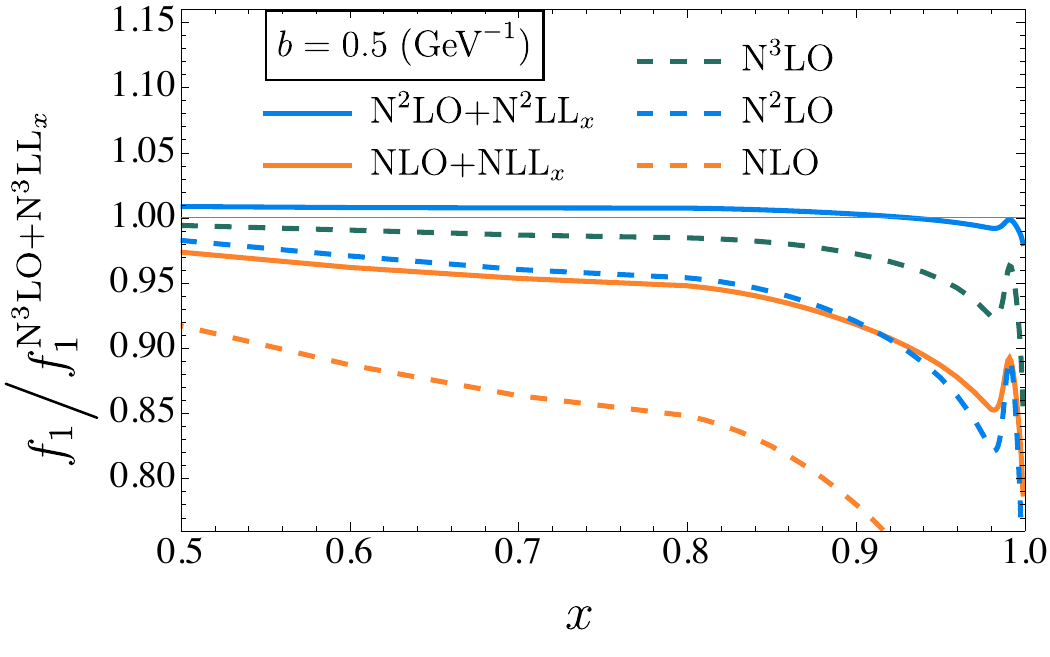}
\includegraphics[width=0.475\linewidth]{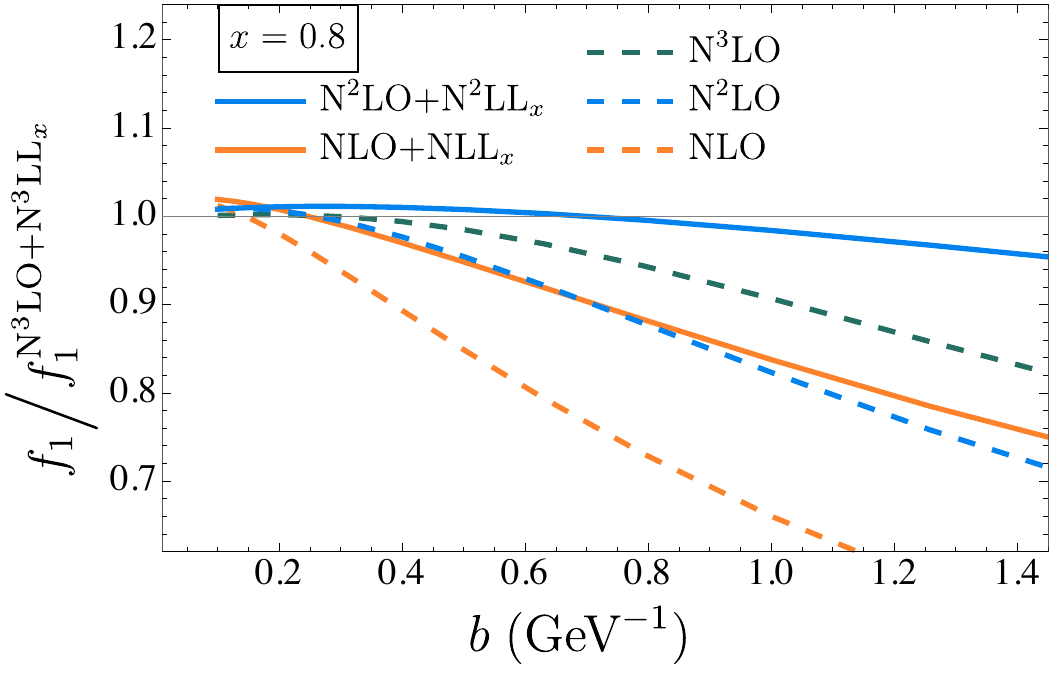}
\caption{\label{fig:uTMDPDF} The comparison of different orders of small-$b$ matching expression ($f_{\text{NP}}=1$) for the optimal unpolarized TMDPDF (for the $u$-quark). The curves in the left and the right panels are results divided by N$^3$LO + N$^3$LL$_x$ computations. }
\end{figure}

In fig.~\ref{fig:uTMDPDF}, we present various curves for the convolution in eq.~(\ref{def:tw2-Convol}) in the case of the unpolarized TMDPDF, considering only the perturbative contribution. The unpolarized PDF is taken from the MSHT20 extraction \cite{Bailey:2020ooq}. Computations performed at a fixed order are shown with dashed lines, while those using the resummed expression \eqref{def:Cresum} are displayed with solid lines. All curves are normalized to the highest perturbative precision available, N$^3$LO+N$^3$LL$_x$. We observe the following behavior:
\begin{itemize}
\item As expected, resummation effects are significant at large $x$ but diminish as $x$ decreases, becoming negligible for $x \lesssim 0.2$. These effects also reduce as $b$ approaches zero due to faster convergence of the perturbative series. The main region where resummation has an impact is $x \gtrsim 0.3$ and $b \gtrsim 0.3\,\, \mathrm{GeV}^{-1}$. In the following, we restrict our focus to this region. Note, that for large-$b$ any perturbative behavior is dominated by non-perturbative effects.
\item The resummed expression exhibits faster convergence with increasing resummation order than the fixed-order results. Specifically, the difference between N$^3$LL$_x$ and N$^2$LL$_x$ is less than 1\%, whereas the difference between N$^3$LO and N$^2$LO ranges from 2\% to 4\% and grows significantly as $x \to 1$.
\item Including resummation effectively approximates higher-order corrections. For example, NLO + NLL$_x$ closely reproduces N$^2$LO, demonstrating that resummation is particularly useful when higher-order corrections are unavailable.
\item The difference between N$^3$LO order and the resummed expression is small, of the order of $1-2$\% for $x<0.8$ and $b<1$ GeV$^{-1}$. This indicates that the perturbative expression is practically saturated at the three-loop order.
\end{itemize}
Therefore, resummation can significantly enhance theoretical predictions in cases where higher-order perturbative terms are unavailable. This is particularly relevant for helicity TMD distributions and, to a lesser extent, transversity TMD distributions. In contrast, for unpolarized TMDPDFs and TMDFFs, where N$^3$LO results are known, resummation introduces only minor modifications.

\subsection{Wandzura-Wilczek part of worm-gear functions}

Worm-gear functions consist of both WW and twist-three terms, which play distinct roles in the large-$x$ resummation. As discussed, all singular contributions in the WW part can be resummed, but these are only logarithmically singular. In contrast, for the twist-three terms, only the leading asymptotic contribution can be resummed. We omit a detailed discussion of the twist-three part for the following reasons: theoretically, these terms are analogous to the Sivers and Boer-Mulders distributions, though with much more complex expressions \cite{Rein:2022odl}; practically, they are difficult to implement, and the corresponding collinear distributions are not yet available. However, we have explicitly verified that the leading large-$x$ singularity matches the predicted one. For these reasons, we focus on the WW part in this section.

As derived in sec.~\ref{sec:WW} the expressions for WW parts of worm-gear functions are
\begin{eqnarray}
g_{1T,f}^{\perp,\text{WW}}(x,b)&=&x\int_x^1 \frac{dy}{y}\sum_{f'}\(
\delta_{ff'}V_f^{\text{WW}}\(\frac{x}{y},b;\mu,\zeta\)+\Delta C^{(g)}_{f\ot f'}\(\frac{x}{y},b;\mu,\zeta\)
\)g_{1,f'}(y),
\\
h_{1L,f}^{\perp,\text{WW}}(x,b)&=&-x^2\int_x^1 \frac{dy}{y^2}
\sum_{f'}\(
\delta_{ff'}V_f^{\text{WW}}\(\frac{x}{y},b;\mu,\zeta\)+\Delta C^{(h)}_{f\ot f'}\(\frac{x}{y},b;\mu,\zeta\)
\)
h_{1,f'}(y),
\end{eqnarray}
where function $V$ is
\begin{eqnarray}
V_f^{\text{WW}}(x,b;\mu,\zeta)=\frac{e^{\overline{\mathcal{E}}_f}}{(1-x)^{\alpha_f}},
\end{eqnarray}
and $\Delta C$ are finite at $x\to1$.  The function $V$ is universal for both cases and consists of universal elements that are known up to N$^3$LO. In contrast, the terms $\Delta C$ are known only at one-loop \cite{Rein:2022odl} and only for the quark channel and read\footnote{
There is a misprint in the sign of the $\sim \mathbf{L}_\mu$ term for $C_{1L,q\ot q}^{\perp,\text{tw2}}$ in ref.\cite{Rein:2022odl}. The corrected expression must read
\begin{eqnarray}
C_{1L,q\ot q}^{\perp,\text{tw2}}&=&
1+a_sC_F\Big[-\mathbf{L}_\mu^2+2\mathbf{L}_\mu\mathbf{l}_\zeta+4 \mathbf{L}_\mu (\ln x -\ln \bar x)-\frac{\pi^2}{6}\Big]+\mathcal{O}(a_s^2).
\end{eqnarray}
It follows directly from the expression for diagrams $A+A^*$ given in appendix B.3 of the same article.
}
\begin{eqnarray}
\Delta C^{(g)}_{q\ot q}&=&1+a_s C_F(2\mathbf{L}_\mu-1)(\bar x+\ln x)+\mathcal{O}(a_s^2),
\\
\Delta C^{(g)}_{q\ot g}&=&1-\frac{a_s}{2} C_F(2\mathbf{L}_\mu-1)(2\bar x+\ln x)+\mathcal{O}(a_s^2),
\\
\Delta C^{(h)}_{q\ot q}&=&1+4a_s C_F \mathbf{L}_\mu \ln x +\mathcal{O}(a_s^2),
\end{eqnarray}
where only diagonal-flavor functions are modified by resummation.

\begin{figure}
\centering
\includegraphics[width=0.48\linewidth]{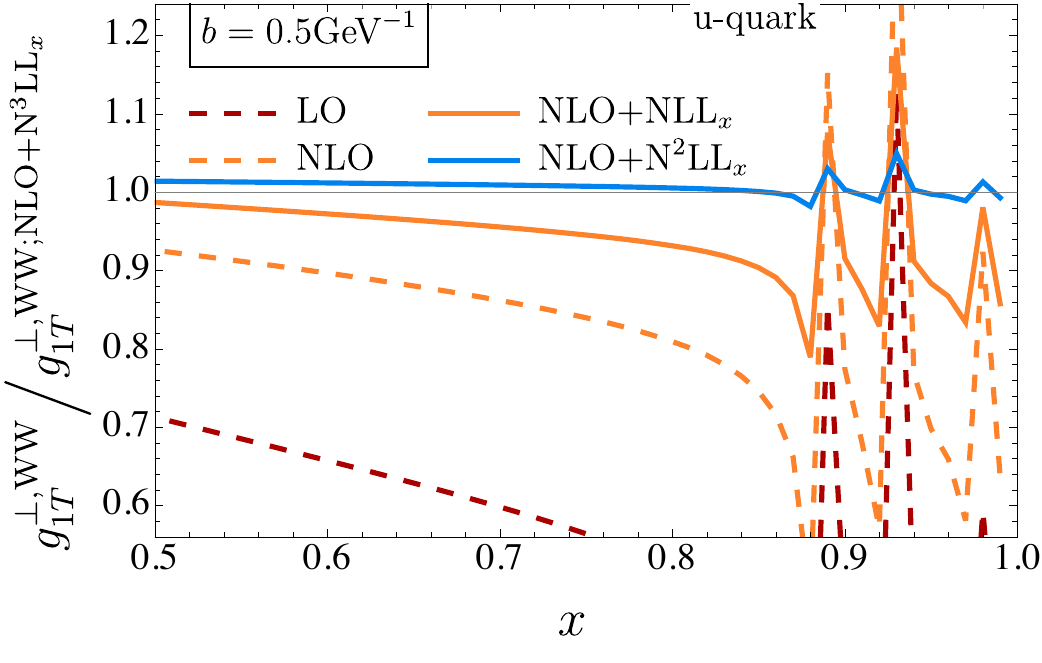}
~~
\includegraphics[width=0.48\linewidth]{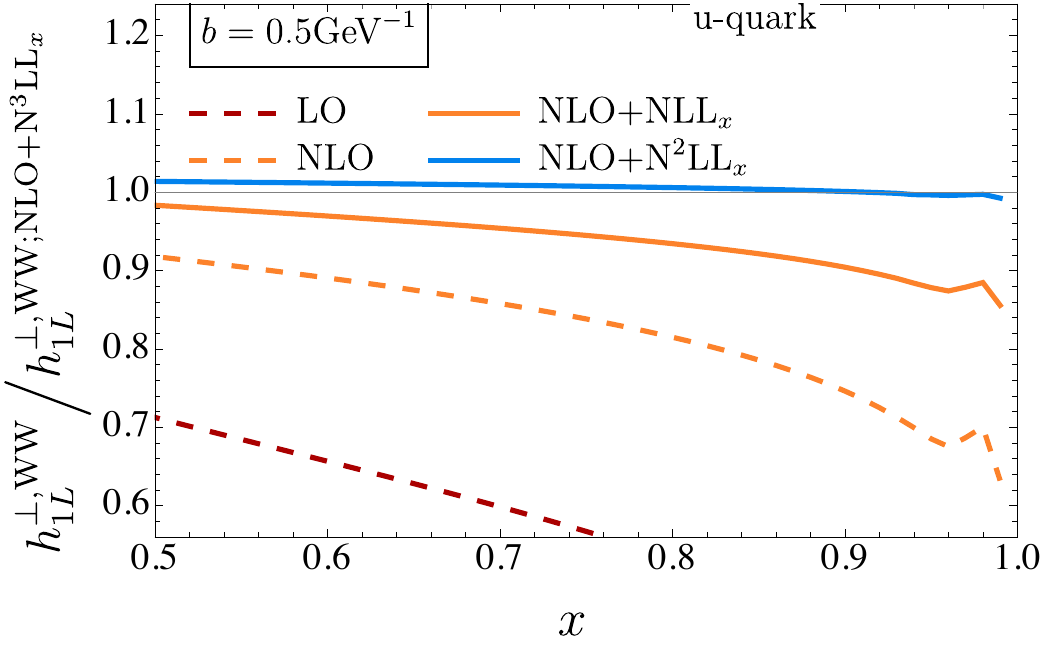}
\caption{\label{fig:wgtTMDPDF} Comparison of different orders of small-$b$ matching expression ($f_{\text{NP}}=1$) for $u$-quark optimal worm-gear-T (left) and worm-gear-L (right) TMDPDFs. The curves are weighted to NLO+N$^3$LL$_x$ expression. The oscillations at large-$x$ are due to oscillations of collinear PDF around zero.}
\end{figure}

In fig.~\ref{fig:wgtTMDPDF}, we compare the resummed and fixed-order convolutions for the worm-gear-T and worm-gear-L distributions. The calculation for $g_{1T}^\perp$ uses $g_1$ from the DSSV19 set \cite{DeFlorian:2019xxt}, while the computation for $h_{1L}^\perp$ is based on $h_1$ from the JAM22 set \cite{Gamberg:2022kdb}. Both distributions exhibit similar behavior. As with the unpolarized TMDPDF, the resummed expression demonstrates perturbative convergence. Since only the NLO part of the coefficient function is available, the modifications introduced by resummation are numerically significant. The N$^3$LL resummation, applied on top of the NLO result, leads to modifications of about 10\% across the entire range of $x$. This effect is primarily due to the fact that both distributions approach zero as $x \to 0$, meaning that the large-$x$ region dominates. It is also worth noting the oscillations in $g_{1T}^\perp$ at very large $x$, which arise from oscillations in the collinear PDF around zero. These oscillations result in non-positive-definite values (albeit very small, $\sim 10^{-6}$), which are amplified in the ratio.

\section{Conclusions}
\label{sec:conclusions}
The matching of TMD distributions to collinear distributions is a crucial aspect of TMD phenomenology. Numerous studies have confirmed that incorporating higher-order perturbative inputs significantly improves the description of experimental data. The large-$x$ resummation discussed in this work serves a similar purpose. It provides means to estimate unknown higher-order contributions to the coefficient function in the large-$x$ regime and enhances the convergence of the perturbative series.

In contrast to previous studies of large-$x$ resummation within TMD factorization, we perform the resummation of large-$x$ terms directly within the TMD distributions. This approach ensures that our results are process-independent and universal, as the resulting expressions for TMD distributions can be applied across a wide range of scenarios, from lattice QCD calculations to high-energy jet production. Furthermore, for the first time, we address the large-$x$ asymptotics of twist-three operators. This includes Wandzura-Wilczek terms, for which resummation can be performed to any order, and pure twist-three terms, for which resummation is achievable only at LL$_x$. Consequently, our resummation expressions encompass all leading-power TMD distributions, including PDFs and FFs, with the exception of the pretzelosity distribution, which is associated with a twist-four operator.

Using the known orders of the anomalous dimension, one can perform large-$x$ resummation up to the N$^3$LL$_x$ level, which is generally higher than the currently known orders of the coefficient functions. For unpolarized TMD distributions, where the coefficient function is known at N$^3$LO, the modifications introduced by large-$x$ resummation are small, indicating excellent perturbative convergence of the series. In contrast, for other cases, only the NLO parts of the coefficient functions are known, except for the transversity TMDPDF, which is known at N$^2$LO. In these cases, the modifications due to large-$x$ resummation are significant. Since these distributions are dominated by their large-$x$ behavior (with their small-$x$ contributions being suppressed), the formulas derived in this work are of substantial practical importance.

\acknowledgments

This work was partially supported by the National Science Foundation under Grants  No.~PHY-2310031, and No.~PHY-2335114 (A.P.), and the U.S. Department of Energy Contract No.~DE-AC05-06OR23177, under which Jefferson Science Associates, LLC operates Jefferson Lab (A.P.).
A.V. is funded by the \textit{Atracci\'on de Talento Investigador} program of the Comunidad de Madrid (Spain) No. 2020-T1/TIC-20204 and \textit{Europa Excelencia} EUR2023-143460, MCIN/AEI/10.13039/501100011033/,  from Spanish Ministerio de Ciencias y Innovaci\'on. O. dR. is supported by the MIU (Ministerio de Universidades, Spain) fellowship FPU20/03110. 
This project is supported by the Spanish Ministerio de Ciencias y Innovaci\'on Grant No. PID2022-136510NB-C31 funded by MCIN/AEI/ 10.13039/501100011033. This project has received funding from the European Union Horizon research Marie Skłodowska-Curie Actions – Staff Exchanges, HORIZON-MSCA-2023-SE-01-101182937-HeI.

\appendix 

\section{Expressions for anomalous dimensions}
\label{app:PT}

There are two perturbative series important for the large-$x$ resummation of TMD distributions. These are rapidity anomalous dimension, $\mathcal{D}_{\text{pt}}$, and the renormalized soft factor, $\mathcal{E}$. In this appendix, we collect expressions necessary for the application of large-$x$ resummation up to N$^3$LL$_x$. We utilize the following notation for the QCD coupling constant and for the logarithms
\begin{eqnarray}
a_s=\frac{g^2}{(4\pi)^2},\qquad \mathbf{L}_\mu=\ln\(\frac{\mu^2 \vec b^2}{4e^{-2\gamma_E}}\),\qquad \mathbf{l}_\zeta=\ln\(\frac{\mu^2}{\zeta}\).
\end{eqnarray}
The QCD beta function is
\begin{eqnarray}
\frac{d a_s}{d\ln \mu^2}=-\beta(\mu)=-a_s^2 \sum_{n=0}^\infty a_s^n\beta_n.
\end{eqnarray}
The series of anomalous dimensions is defined as
\begin{eqnarray}
\Gamma^f_{\text{cusp}}(\mu)=a_s\sum_{n=0}^\infty a_s^n \Gamma^f_n,\qquad
\gamma^f_S(\mu)=a_s\sum_{n=0}^\infty a_s^n \gamma^{S,f}_n,
\end{eqnarray}
where $\Gamma^f_{\text{cusp}}$ is the cusp anomalous dimension, and $\gamma^f_S$ is the soft anomalous dimension, and $f$ stands for either quarks or gluons. The index $f$ indicates quark or gluon. Note, that both anomalous dimensions satisfy Casimir scaling $\Gamma^q=C_F/C_A\Gamma^g$ up to three loops, and the generalized Casimir scaling at four-loops \cite{Moch:2017uml}. The compendium of expressions for these anomalous dimensions up to four loops can be found, for example, in ref.~\cite{Moult:2022xzt}.

\subsection*{Rapidity anomalous dimension}

The perturbative expansion of the rapidity anomalous dimension reads
\begin{eqnarray}
\label{eq:CS}
\mathcal{D}^f_{\text{pt}}(b;\mu)&=&\sum_{n=1}^\infty \sum_{k=0}^n a_s^n \mathbf{L}_\mu^k d_f^{(n,k)}.
\end{eqnarray}
Coefficients $d^{(n,k)}$ with $k\neq 0$ are determined by the renormalization group equation
\begin{equation}
\frac{d\mathcal{D}_{\text{pt}}(b;\mu)}{d\ln\mu^2}=\frac{\Gamma_{\text{cusp}}^f(\mu)}{2},
\end{equation}
and read
\begin{align}
\nn 
d^{(1,1)}=&\frac{\Gamma_0}{2},
&
d^{(2,1)}=&\frac{\Gamma_1}{2},
\\\nn 
d^{(2,2)}=&\frac{\beta_0\Gamma_0}{4},
&
d^{(3,1)}=&\frac{\Gamma_2}{2}+2\beta_0d^{(2,0)} 
\\
d^{(3,2)}=& \frac{\beta_1\Gamma_0+2\beta_0\Gamma_1}{4}
& 
d^{(3,3)}=&\frac{\beta_0^2\Gamma_0}{6},
\\\nn 
d^{(4,1)}=&
\frac{\Gamma_3}{2}+2\beta_1d^{(2,0)}+3\beta_{0} d^{(3,0)}
&
d^{(4,2)}=\,&
\frac{\beta_2\Gamma_0+2\beta_1\Gamma_1+3\beta_0\Gamma_2}{4}+3\beta_0^2d^{(2,0)}
\\\nn
d^{(4,3)}=\,&
\frac{5\beta_0\beta_1\Gamma_0+6\beta_0^2\Gamma_1}{12}
&
d^{(4,4)}=\,&\frac{\beta_0^3\Gamma_0}{8},
\end{align}
where $\Gamma_n$ are coefficients of the cusp-anomalous dimension of the corresponding flavor. The expression for $d^{(n,0)}$ are known up to four loops \cite{Echevarria:2016scs, Vladimirov:2016dll, Duhr:2022yyp, Moult:2022xzt}, and read
\begin{align}
 d_q^{(1,0)}=&\,0,
\\ 
d_q^{(2,0)}=&C_F\[\,C_A\left(\frac{404}{27}-14 \zeta_3\right)-\frac{56}{27}N_f\],
\\ 
d_q^{(3,0)}=&C_F\Big[\,C_A^2\left(\frac{297029}{1458}-\frac{3196}{81}\zeta_2-\frac{6164}{27}\zeta_3-\frac{77}{3}\zeta_4+\frac{88}{3}\zeta_2\zeta_3+96\zeta_5\right)
\\\nn&
+C_AN_f\left(-\frac{31313}{729}+\frac{412}{81}\zeta_2+\frac{452}{27}\zeta_3-\frac{10}{3}\zeta_4\right)
+C_FN_f\left(-\frac{1711}{54}+\frac{152}{9}\zeta_3+8\zeta_4\right)
\\\nn&
+N_f^2\left(\frac{928}{729}+\frac{16}{9}\zeta_3\right)\Big].
\end{align}
The expression for the $d^{(4,0)}$ is rather lengthy and can be found in refs.~\cite{Moult:2022xzt, Duhr:2022yyp}. The numerical values of rapidity anomalous dimension up to 16 digits are
\begin{align}
\nn d^{(1,0)}_q=&0,
\\
d^{(2,0)}_q=&-7.463334725085428 - 2.765432098765432 N_f,
\\
\nn d^{(3,0)}_q=&-70.06800923663323 - 77.12861616470005 N_f + 4.546622307260785 N_f^2,
\\
\nn d^{(4,0)}_q=&350.8342523981021 - 2428.14 N_f + 378.3057617542652 N_f^2 - 
 8.071924959941491 N_f^3.
\end{align}
The expressions for gluon rapidity anomalous dimensions $d^{(n,0)}_g$ can be obtained by Casimir scaling 
\begin{equation}
d^{(n,k)}_g=\frac{C_A}{C_F}d^{(n,k)}_q=\frac{9}{4}d^{(n,k)}_q,
\end{equation}
except for $d_g^{(4,0)}$, which obeys a generalized Casimir scaling. The numerical value of $d^{(4,0)}_g$ is
\begin{eqnarray}
d^{(4,0)}_g&=&
-333.769693860101 - 5506.38 N_f + 851.1879639470966 N_f^2 
- 18.16183115986835 N_f^3. \nonumber \\
\end{eqnarray}

\subsection*{Renormalized soft factor}

The expression for $\mathcal{E}$ at general TMD scales can be read directly from refs.~\cite{Ebert:2020yqt, Luo:2020epw} up to three-loop order. Here, we present a version decomposed over universal anomalous dimensions. 

The renormalized soft factor $\mathcal{E}$ for general scales obeys the following renormalization group equations
\begin{eqnarray}
\frac{d\mathcal{E}^f(b;\mu,\zeta)}{d\ln \zeta}=-\mathcal{D}_{\text{pt}}^f(b;\mu),\qquad 
\frac{d\mathcal{E}^f(b;\mu,\zeta)}{d\ln \mu^2}=\frac{\Gamma_{\text{cusp}}^f(\mu)}{2}\mathbf{l}_\zeta-\gamma_S^f(\mu),
\end{eqnarray}
where $\gamma_S$ is the soft anomalous dimension \cite{Moch:2004pa, vonManteuffel:2020vjv, Das:2019btv}. The signs and factors $2$ are defined in accordance with the definition of the soft anomalous dimension $\gamma_S$ as given in ref.~\cite{Moult:2022xzt}. It follows that 
\begin{eqnarray}
\mathcal{E}^f(b;\mu,\zeta)=\mathcal{D}^f_{\text{pt}}(b;\mu)\mathbf{l}_\zeta+\sum_{n=1}^\infty \sum_{k=0}^{n+1} a_s^n \mathbf{L}_\mu^k e^{(n,k)}_f,
\end{eqnarray}
with
\begin{align}
\nn 
e^{(1,1)}=&0,
&
e^{(1,2)}=&-\frac{\Gamma_0}{4},
\\\nn 
e^{(2,2)}=&-\frac{\Gamma_1}{4},
&
e^{(2,1)}=&-\gamma^S_1-d^{(2,0)}+\beta_0 e^{(1,0)},
\\\nn 
e^{(2,3)}=&-\frac{\beta_0\Gamma_0}{6},
&
e^{(3,1)}=&-\gamma^S_2-d^{(3,0)}+\beta_1 e^{(1,0)}+2\beta_0e^{(2,0)},
\\
e^{(3,2)}=&-\frac{\Gamma_2+4\beta_0\gamma_1^S+8\beta_0d^{(2,0)}}{4}+\beta_0^2e^{(1,0)},
&
e^{(3,3)}=&-\frac{\beta_1\Gamma_0+2\beta_0\Gamma_1}{6},
\\\nn 
e^{(3,4)}=&-\frac{\beta_0^2\Gamma_0}{8},
&
e^{(4,1)}=&-\gamma^S_3-d^{(4,0)}+\beta_2e^{(1,0)}+2\beta_1e^{(2,0)}+3\beta_0 e^{(3,0)},
\end{align}
\begin{align}\nn
e^{(4,2)}=&
-\frac{\Gamma_3+4\beta_1\gamma_1^S+6\beta_0\gamma_2^S+8\beta_1d^{(2,0)}+12\beta_0d^{(3,0)}}{4}
+\frac{5}{2}\beta_0\beta_1 e^{(1,0)}+3\beta_0^2e^{(2,0)},
\phantom{111111111111111111}
\\\nn 
e^{(4,3)}=&-\frac{\beta_2\Gamma_0+2\beta_1\Gamma_1+3\beta_0\Gamma_2+6\beta_0^2\gamma_1^S+18\beta_0^2d^{(2,0)}}{6}+ \beta_0^3e^{(1,0)},
\end{align}
\begin{align}\nn
e^{(4,4)}=&-\frac{5\beta_0\beta_1\Gamma_0+6\beta_0^2\Gamma_1}{16},
\phantom{111111111111111}
&
e^{(4,5)}=&-\frac{\beta_0^3\Gamma_0}{10},
\phantom{1111111111111111111!!111111}
\end{align}
where coefficients $\Gamma$, $\gamma$ and $d$ should be taken for the  corresponding flavor. To derive these expressions we have used that $\gamma_0^S=0$. The boundary values for the quark case are
\begin{eqnarray}
e_q^{(1,0)}&=&-C_F\zeta_2,
\\
e_q^{(2,0)}&=&C_F\[C_A\(\frac{1214}{81}-\frac{67}{6}\zeta_2-\frac{77}{9}\zeta_3+5\zeta_4\)
+N_f\(-\frac{164}{81}+\frac{5}{3}\zeta_2+\frac{14}{9}\zeta_3\)\]
\\\nn
e_q^{(3,0)}&=&C_F\Big[
C_A^2\Big(\frac{5211949}{26244}-\frac{297481}{1458}\zeta_2-\frac{75566}{243}\zeta_3+\frac{3649}{54}\zeta_4+\frac{902}{9}\zeta_5+\frac{550}{9}\zeta_2\zeta_3 -\frac{1543}{27}\zeta_6
\\\nn &&~
+\frac{464}{9}\zeta_3^2\Big)
+C_AN_f\Big(-\frac{412765}{13122}+\frac{37265}{729}\zeta_2+\frac{4076}{81}\zeta_3-\frac{208}{27}\zeta_4-\frac{92}{3}\zeta_5+\frac{20}{9}\zeta_2\zeta_3\Big)
\\\nn &&~
+C_FN_f\Big(-\frac{42727}{972}+\frac{275}{18}\zeta_2+\frac{1744}{81}\zeta_3+\frac{76}{9}\zeta_4+\frac{112}{9}\zeta_5-\frac{40}{3}\zeta_2\zeta_3\Big)
\\ &&~
+N_f^2\Big(-\frac{128}{6561}-\frac{68}{27}\zeta_2-\frac{280}{243}\zeta_3-\frac{22}{27}\zeta_4\Big)\Big].
\end{eqnarray}
The corresponding numerical values for $e^{(n,0)}$ up to 16 digits are
\begin{eqnarray*}
e^{(1,0)}&=&-2.193245422464302,
\\
e^{(2,0)}&=&-33.01369815806463 + 3.448975618479332 N_f,
\\
e^{(3,0)}&=&-2358.961879817530 + 314.3621232240640 N_f - 8.572380372078722 N_f^2.
\end{eqnarray*}
The expressions for the gluon coefficients $e_g^{(n,k)}$ can be obtained by Casimir scaling
$$e_g^{(n,k)}=\frac{C_A}{C_F}e_q^{(n,k)}=\frac{9}{4}e_q^{(n,k)}.$$
The expression for $e^{(4,0)}$ is presently unknown.

\subsection*{Renormalized soft factor in the $\zeta$-prescription}

The expression for $\mathcal{E}$ in the $\zeta$-prescription can be obtained by substituting the perturbative expression for $\mathbf{l}_\zeta$ (see appendix A in ref.~\cite{Scimemi:2019cmh}). In the $\zeta$-prescription the $\mathcal{E}$ has a single-logarithm perturbative expansion
\begin{eqnarray}
\mathcal{E}(b;\mu)&=&\sum_{n=0}^\infty \sum_{k=0}^{n} a_s^n \mathbf{L}_\mu^k \mathfrak{e}^{(n,k)}_f,
\end{eqnarray}
and satisfies the renormalization group equation 
\begin{eqnarray}\label{app:gamma-delta}
\frac{d\mathcal{E}^f(b;\mu)}{d\ln \mu^2}=-\gamma^f_\Delta=-\gamma_S^f(\mu)+\frac{\gamma^f_{V}(\mu)}{2},
\end{eqnarray}
where $\gamma^f_{V}$ is the TMD anomalous dimension in the traditional definition (see for instance refs.~\cite{Scimemi:2019cmh, Moos:2023yfa}). The expression for $\gamma_V$ is known up to four-loops \cite{Agarwal:2021zft}. Consequently, the terms $\mathfrak{e}^{(n,k)}$ with $k>0$ are
\begin{align}
\nn 
\mathfrak{e}^{(1,1)}=&-\gamma^\Delta_0,
&
\mathfrak{e}^{(2,1)}=&-\gamma^\Delta_1+\beta_0 \mathfrak{e}^{(1,0)},
\\\nn 
\mathfrak{e}^{(2,2)}=&-\frac{\beta_0\gamma^\Delta_0}{2},
&
\mathfrak{e}^{(3,1)}=&-\gamma^\Delta_2+\beta_1\mathfrak{e}^{(1,0)}+2\beta_0 \mathfrak{e}^{(2,0)},
\\ 
\mathfrak{e}^{(3,2)}=&-\frac{\beta_1\gamma^\Delta_0+2\beta_0\gamma^\Delta_1}{2}+\beta_0^2\mathfrak{e}^{(1,0)},
&
\mathfrak{e}^{(3,3)}=&-\frac{\beta_0^2\gamma^\Delta_0}{3},
\\\nn 
\mathfrak{e}^{(4,1)}=&-\gamma^\Delta_3+\beta_2\mathfrak{e}^{(1,0)}+2\beta_1\mathfrak{e}^{(2,0)}+3\beta_0\mathfrak{e}^{(3,0)},
&
\mathfrak{e}^{(4,2)}=&-\frac{\beta_2\gamma^\Delta_0+2\beta_1\gamma^\Delta_1+3\beta_0\gamma^\Delta_2}{2}
\\\nn & & & +\frac{5}{2}\beta_0\beta_1\mathfrak{e}^{(1,0)}+3\beta_0^2\mathfrak{e}^{(2,0)}
,
\\\nn 
\mathfrak{e}^{(4,3)}=&-\frac{5\beta_0\beta_1\gamma^\Delta_0+6\beta_0^2\gamma^\Delta_1}{6}+\beta_0^3\mathfrak{e}^{(1,0)},
&
\mathfrak{e}^{(4,4)}=&-\frac{\beta_0^3\gamma^\Delta_0}{4}
,
\end{align}
where
\begin{eqnarray}
\gamma_\Delta^f=a_s\sum_{n=0}^\infty a_s^n \gamma_n^{\Delta}.
\end{eqnarray}
These anomalous dimensions up to three-loops (for quark flavor) are
\begin{eqnarray}
\gamma_0^{\Delta}&=&3C_F,
\\
\gamma_1^{\Delta}&=&C_F\Big[C_F\(\frac{3}{2}-12\zeta_2+24\zeta_3\)+C_A\(\frac{51}{18}+\frac{44}{3}\zeta_2-12\zeta_3\)+N_f\(-\frac{1}{3}-\frac{8}{3}\zeta_2\)\Big],
\\
\gamma_2^{\Delta}&=&C_F\Big[C_F^2\Big(
\frac{29}{2}+18\zeta_2+68\zeta_3+144\zeta_4-240\zeta_5-32\zeta_2\zeta_3\Big)
+C_A^2\Big(-\frac{1657}{36}+\frac{4496}{27}\zeta_2
\\\nn &&~
-\frac{1552}{9}\zeta_3-5\zeta_4+40\zeta_5\Big)
+C_FC_A\Big(\frac{151}{4}-\frac{410}{3}\zeta_2+\frac{844}{3}\zeta_3-\frac{494}{3}\zeta_4+120\zeta_5+\frac{48}{3}\zeta_2\zeta_3\Big)
\\\nn &&~
+C_FN_f\Big(-23+\frac{20}{3}\zeta_2-\frac{136}{3}\zeta_3+\frac{116}{3}\zeta_4\Big)
+C_AN_f\Big(20-\frac{1336}{27}\zeta_2+\frac{200}{9}\zeta_3+2\zeta_4\Big)
\\\nn &&~
+N_f^2\Big(-\frac{17}{9}+\frac{80}{27}\zeta_2-\frac{16}{9}\zeta_3\Big)\Big].
\end{eqnarray}
The gluon case can be obtained by Casimir scaling. The fourth-loop expression can be combined using the definition (\ref{app:gamma-delta}) and expressions in refs.~\cite{Agarwal:2021zft, Moult:2022xzt}. The boundary values for the quark case are
\begin{eqnarray}
\mathfrak{e}_q^{(1,0)}&=&-C_F\zeta_2,
\\
\mathfrak{e}_q^{(2,0)}&=&C_F\[C_A\(-\frac{604}{81}-\frac{67}{6}\zeta_2+\frac{112}{9}\zeta_3+5\zeta_4\)
+N_f\(\frac{88}{81}+\frac{5}{3}\zeta_2+\frac{14}{9}\zeta_3\)\]
\\\nn
\mathfrak{e}_q^{(3,0)}&=&C_F\Big[
C_A^2\Big(-\frac{224035}{13122}-\frac{396625}{1458}\zeta_2+\frac{43493}{486}\zeta_3+\frac{2864}{27}\zeta_4-\frac{394}{9}\zeta_5+\frac{1225}{9}\zeta_2\zeta_3 -\frac{1543}{27}\zeta_6
\\\nn &&~
-\frac{733}{9}\zeta_3^2\Big)
+C_FC_A(202-189\zeta_3)\Big(-\frac{1}{18}+\frac{4}{9}\zeta_2-\frac{8}{9}\zeta_3\Big)
+C_AN_f\Big(\frac{38215}{6561}+\frac{55463}{729}\zeta_2
\\\nn &&~
+\frac{1559}{81}\zeta_3-\frac{73}{27}\zeta_4-\frac{92}{3}\zeta_5-\frac{106}{9}\zeta_2\zeta_3\Big)
+C_FN_f\Big(\frac{2491}{486}+\frac{17}{6}\zeta_2+\frac{1708}{81}\zeta_3-\frac{32}{9}\zeta_4
+\frac{112}{9}\zeta_5
\\ &&~
-\frac{40}{3}\zeta_2\zeta_3\Big)
+N_f^2\Big(\frac{700}{6561}-\frac{124}{27}\zeta_2-\frac{928}{243}\zeta_3-\frac{22}{27}\zeta_4\Big)\Big].
\end{eqnarray}
The corresponding numerical values for $\mathfrak{e}^{(n,0)}$ up to 16 digits are
\begin{eqnarray*}
\mathfrak{e}^{(1,0)}&=&-2.193245422464302,
\\
\mathfrak{e}^{(2,0)}&=&-21.81869607043648 + 7.597123766627481 N_f,
\\
\mathfrak{e}^{(3,0)}&=&-2278.243647000777 + 416.0684169124957 N_f - 17.22704690855201 N_f^2.
\end{eqnarray*}
The expressions for gluon coefficients $\mathfrak{e}_g^{(n,k)}$ can be obtained by Casimir scaling. The expression for $\mathfrak{e}^{(4,0)}$ is presently unknown.

\section{Comparison with prior resummation results}
\label{app:ComparisonKang}

Previous studies \cite{Kulesza:2002rh, Kulesza:2003wn, Lustermans:2016nvk, Procura:2018zpn, Kang:2022nft} have performed resummation in the large-$x$ regime in the joint resummation formalism for the unpolarised case. In this Appendix, we compare with their results in order to demonstrate the consistency of our work. Usually, resummation formulas are presented in Mellin space. For instance, consider eq.~(2) from Ref.\cite{Kang:2022nft} for the OPE in the large-$N$ regime which reads
\begin{eqnarray}\label{ResummMellin}
\lim_{N\to\infty}\mathbf{M}_N[F_{f\ot h}(x,b;\mu,\zeta)]=\Tilde{S}_c^f(b;\mu,\zeta_N) \mathbf{M}_N[f_{f\ot h}\(x;\mu\)]+\mathcal{O}(\vec b^2),
\end{eqnarray}
where $\zeta_N=\zeta/\Bar{N}^2$ and $\Tilde{S}_c$ is the collinear-soft function (in the nomenclature of ref.~\cite{Kang:2022nft}). To align this with eq.~\eqref{MN=e^D}, we must verify the following relation:
\begin{equation}\label{eq:RelationToVerify}
    \Tilde{S}_c^f(b;\mu,\zeta_N)=\exp\(2\mathcal{D}^f_{p.t.}(b;\mu)\ln \bar N+\mathcal{E}^f(b;\mu,\zeta)\)\,.
\end{equation}
If we introduce a perturbative expansion of the collinear-soft function
\begin{eqnarray}\label{eq:CollinearSoftFunction}
\Tilde{S}_c^f(b;\mu,\zeta_N)&=&\sum_{n=1}^\infty a^n_s(\mu) S^f_n(b;\mu,\zeta_N)\,,
\end{eqnarray}
then, at one loop order, we must verify that
\begin{equation}
    S_1^f=2 D^f_1\ln \bar N+E_1^f\,.
\end{equation}
Using the expressions for $D_1^q$ and $E_1^q$ provided in Appendix \ref{app:PT}, we find:
\begin{equation}
    S_1^q=C_F\[-\mathbf{L}_\mu^2+2 \mathbf{L}_\mu\ln\(\frac{\mu^2}{\zeta_N}\)-\zeta_2\]\,.
\end{equation}
which matches eq.~(A.7) from Ref.\cite{Kang:2022nft}. Moreover, the unsubtracted collinear-soft function is defined as
\begin{equation}\label{eq:Collinear-Soft}
    \Tilde{S}_c^f(b;\mu,\zeta_N)=\Tilde{S}_c^{f,\text{unsub}}(b;\mu,\zeta_N/\nu^2)\sqrt{S^f(b;\mu,\nu)}\,,
\end{equation}
where $S^f$ is the standard TMD soft function and $\nu$ is the rapidity scale. The dependence on $\nu$ cancels between the terms on the right-hand side. By comparing to Ref.\cite{Kang:2022nft}, we confirm the following relations up to three-loop order:
\begin{eqnarray}
    &&\Tilde{S}_c^{f,\text{unsub}}(b;\mu,\zeta_N/\mu^2)=\exp\(\mathcal{D}^f_{p.t.}(b;\mu)\ln\(\frac{\mu^2}{\zeta_N}\)\)\,,\\
    &&\sqrt{S^f(b;\mu,\mu)}=\exp\(\mathcal{E}^f(b;\mu,\zeta)-\mathcal{D}^f_{\text{pt}}(b;\mu)\mathbf{l}_\zeta\)\,.
\end{eqnarray}
Substituting these into eq.~\eqref{eq:Collinear-Soft}, we validate the consistency of eq.~\eqref{eq:RelationToVerify}, which establishes the agreement between our findings and previous results in the literature.

\section{Transformation from Mellin space to the momentum-fraction space }
\label{app:MellinToMomentum}
In this section, we derive eq.~(\ref{eq:MellinToMomentumFraction}), which is employed to transform expressions from Mellin space to momentum fraction space. The derivation starts with the Mellin transform of the plus distribution:
\begin{equation}
    \mathbf{M}_N\[\frac{1}{(1-x)^{1+\alpha}_+}\]=\int_0^1dx\frac{x^{N-1}-1}{(1-x)^{1+\alpha}}=\frac{1}{\alpha}+\frac{\Gamma(-\alpha)\Gamma(N)}{\Gamma(N-\alpha)}\quad\quad(\alpha<1)\,.
\end{equation}
Applying Stirling's approximation we obtain the asymptotic behavior of the gamma function for $N\rightarrow\infty$, which reads
\begin{equation}
    \Gamma(N-\alpha)\simeq\Gamma(N)N^{-\alpha}\,.
\end{equation}
Then, evaluating this limit in Mellin space,  which corresponds to the large-$x$ asymptotics in momentum space, we arrive at:
\begin{equation}
    \lim_{N\rightarrow\infty}\mathbf{M}_N\[\frac{1}{(1-x)^{1+\alpha}_+}\]=\frac{1}{\alpha}+N^{\alpha}\Gamma(-\alpha)\,.
\end{equation}
Next, multiplying by a factor that is constant in $N$, yields
\begin{equation}
    \lim_{N\rightarrow\infty}\mathbf{M}_N\[\frac{e^{\alpha\gamma_E}}{\Gamma(-\alpha)}\frac{1}{(1-x)^{1+\alpha}_+}\]=\frac{e^{\alpha\gamma_E}}{\alpha\Gamma(-\alpha)}+e^{\alpha\ln\bar{N}}\,.
\end{equation}
Here, the first term on the right-hand side is independent of $N$. Using eq.~(\ref{eq:MellinDelta}), this term can be transferred to the left-hand side,
\begin{align}
    \frac{e^{\alpha\gamma_E}}{\Gamma(1-\alpha)}\lim_{N\rightarrow\infty}\mathbf{M}_N\[\delta(1-x)-\frac{\alpha}{(1-x)^{1+\alpha}_+}\]=e^{\alpha\ln\bar{N}}\,.
\end{align}
To proceed, we express the common exponential factor using the Taylor series expansion for the natural logarithm of the gamma function
\begin{equation}
    \ln\Gamma(1-\alpha)=\gamma_E\alpha+\sum_{k=2}^\infty\frac{\zeta_k}{k}\alpha^k\quad\quad (|\alpha|<1)\,.
\end{equation}
Finally, multiplying by a constant factor $e^\beta$, we arrive at the relation shown in eq.~(\ref{eq:MellinToMomentumFraction})
\begin{align}
  e^{\beta-\sum_{k=2}^\infty\frac{\zeta_k}{k}\alpha^k}\lim_{N\rightarrow\infty}\mathbf{M}_N\[\delta(1-x)-\frac{\alpha}{(1-x)^{1+\alpha}_+}{\Gamma(-\alpha)}\]=e^{\alpha\ln\bar{N}}\,.
\end{align}

\bibliography{bibFILE}
\end{document}